\documentclass[aps,prb,twocolumn,10pt,superscriptaddress,longbibliography, nofootinbib]{revtex4-2}

\usepackage{mathrsfs}
\usepackage{tabularx}
\usepackage{slashed}
\usepackage{amsmath}
\usepackage{amsfonts}
\usepackage{bm}
\usepackage{soul}
\usepackage{hyperref}
\usepackage{graphicx,color}

\usepackage[dvipsnames]{xcolor} 
\usepackage{natbib}
\usepackage{braket}
\usepackage{tikz}
\usepackage[compat=1.1.0]{tikz-feynman}
\usetikzlibrary{shapes.geometric}
\usetikzlibrary{fit}
\usetikzlibrary{decorations.markings}
\usepackage[caption=false]{subfig}
\usepackage{orcidlink}

\tikzset{
  half solid/.style={
    solid,
    thick
  },
  half dashed/.style={
    dashed,
    thick
  }
}

\tikzset{
  every edge/.style={line cap=butt line join=miter}
}
\tikzset{
  every path/.append style={line cap=butt line join=miter}
}

\tikzset{
  arrow only/.style={
    draw=none,
    postaction={
      decorate,
      decoration={
        markings,
        mark=at position 0.5 with {\arrow[scale=1.6]{>}}
      }
    }
  }
}

\AtBeginDocument{%
   ~\newwrite\bibnotes
   ~\def\bibnotesext{notes.bib}
   ~\immediate\openout\bibnotes=\jobname\bibnotesext
   ~\immediate\write\bibnotes{@CONTROL{REVTEX41Control}}
   ~\immediate\write\bibnotes{@CONTROL{%
    apsrev42Control,author="08",editor="1",pages="1",title="0",year="1"}}
    ~\if@filesw
    ~\immediate\write\@auxout{\string\citation{apsrev42Control}}%
   ~\fi
}%

\begin{document}
\usetikzlibrary{shapes,snakes}
\title{Hall conductance in a weakly time-reversal invariant open system}
\author{Alexander Fagerlund\orcidlink{0009-0009-3196-0286}}
 \email{alexander.fagerlund@fysik.su.se}
 \affiliation{%
Department of Physics, Stockholm University, AlbaNova University Center, SE-106 91 Stockholm, Sweden
}
\author{Christopher Ekman\orcidlink{https://orcid.org/0009-0003-6426-2376}}%
 \email{christopher.ekman@fysik.su.se}
 \affiliation{%
Department of Physics, Stockholm University, AlbaNova University Center, SE-106 91 Stockholm, Sweden
}
 \author{Rodrigo Arouca\orcidlink{0000-0003-4214-1437}}
 \email{rodrigo-arouca@cbpf.br}
\affiliation{%
Brazilian Center For Research in Physics, Rua Doutor Xavier Sigaud 150, Rio de Janeiro, 22290-180, Brazil
}%
\affiliation{%
Department of Physics and Astronomy, Uppsala University, Box 516, S-751 20 Uppsala, Sweden
}%

\begin{abstract}
The quantum Hall effect and the quantum anomalous Hall effect both require time-reversal invariance to be broken. We show that non-equilibrium effects can cause Hall physics to arise even when the system is weakly time-reversal symmetric and no magnetic field is applied. In our model, this occurs due to a fermionic subsystem breaking time-reversal invariance even if the system as a whole does not. The fermions receive a TRI-breaking self-energy, caused by interactions with bosonic degrees of freedom in the system and with an external reservoir. As a result, the fermions develop a non-quantized Hall conductance. We demonstrate that, unlike in the equilibrium case, the presence of a mass term is insufficient for the Hall conductance to appear, and wave-function renormalization effects have to be included. 
\end{abstract}

\maketitle

\section{Introduction}
Since the discovery of the quantum Hall effect \cite{klitzing1980qhe,tsui1982fqhe}, quantum Hall systems have repeatedly been shown to host rich and surprising physics. This includes the famously quantized conductivity, but also topological order, experimentally detected Abelian anyons, and a description in terms of topological field theory \cite{wilczek1982fractional,thouless1982quantized,halperin1984statistics,arovas1984statistics,girvin1987order,zhang1989eft,goldman1995tunneling,saminadayar1997charge,nakamura2020observation}. These systems also have interesting bulk-boundary correspondences, e.g. between the spins of anyons in the bulk and spins induced on the edge \cite{comparin2022spin,nardin2023ssr,fagerlund2025spin,fagerlund2025bosonic}. In addition, some quantum Hall systems are predicted to host non-Abelian anyons with rich braiding behavior, which could in principle be used for topological quantum computing \cite{moore1991nonabelions,kitaev2003computation,nayak2008nonabelian}. 

Understanding the fundamentals in quantum Hall physics has led to a wider exploration of topological phases of matter. In particular, the understanding of the role of symmetries has allowed for a systematic classification of non-interacting fermionic topological phases in the Altland-Zirnbauer classification \cite{altland1997nonstandard, schnyder2008classification,kitaev2009table,ryu2010topological}, and many interesting efforts have been made to understand symmetry-protected topological phases for interacting systems \cite{Wen2017RMP}. Although the Altland-Zirnbauer classification was originally developed to understand the role of time-reversal, particle-hole, and chiral symmetry, it is considerably extended when considering crystalline symmetries \cite{fu2011topological, slager2013space} and driven systems \cite{Kitagawa2010PRB, Lindner2011NatPhys}. 

More recently, notions of topology have been extended to systems described by non-Hermitian operators \cite{Ashida2020AP, Bergholtz2021RMP,Okuma2023ARCMP}. Non-Hermitian matrices appear naturally in open quantum systems\cite{Feshbach1962AnnPhys,Dalibard1992PRL,Dum1992PRA,BreuerPetruccione2002,Prosen2008NJP,Rotter2009JPA,Ekman2024PRR}, interacting systems where quasiparticles have a finite lifetime \cite{Kozii2017PRB,Zyuzin2018PRB,Crippa2023PRL,Kozii_2024}, and classical analogues\cite{Helbig2020NatPhys,Weidemann2020Science, ghatak2020}. For the classification of non-Hermitian Hamiltonians, the set of symmetry classes is extended\footnote{Since conjugation and transposition of a non-Hermitian matrix are not equivalent.} \cite{Gong2018PRX, Kawabata2019PRX}, and there is topology associated with the complex energy levels \cite{YaoWang2018PRL, LeeThomale2019PRB, lin_topological_2023, yang2024rpp}. In more complete approaches to open systems, the role of symmetry becomes more subtle due to the distinction between strong and weak symmetries \cite{buca_2012,altland_2021,sa_symmetry_2023}, which does not have an analogue in closed systems.

The connection between discrete symmetries and topology has allowed the integer-quantized Hall conductivity to be understood as a consequence of broken time-reversal invariance (TRI). For the integer quantum Hall effect (IQHE), one can verify from the tenfold way (see Refs. \cite{schnyder2008classification,kitaev2009table,ryu2010topological}) that there is no non-interacting system in two spatial dimensions which has a $\mathbb{Z}$ topological invariant while also possessing time reversal symmetry. Since this invariant is the Chern number responsible for the quantized conductance, a system in equilibrium at zero temperature has to break TRI to display the IQHE. 

In practice, the TR breaking in quantum Hall systems is caused by a strong magnetic field. Although related phenomena -- such as the quantum anomalous Hall effect and Chern insulators -- can possess Hall physics without a magnetic field, they still rely on the breaking of time-reversal invariance \cite{haldane1988aqhe,chang2023colloquium,Liu2024chern}. In the low-energy continuum description, the breaking of TRI is manifest through the presence of a Dirac mass term. The result is a finite Hall conductance, which becomes half-quantized in the limit of vanishing mass. This is known as the parity anomaly \cite{niemi_axial-anomaly-induced_1983, redlich_gauge_1984,alvarez-gaume_gravitational_1984, haldane1988aqhe, arouca2022anomalies}.
 
In the present work, we introduce a system where Hall physics appears \textit{despite} time-reversal invariance remaining unbroken, in the sense of weak symmetries. Our system consists of fermions and bosons in a 2D sample. Although the dimensionality is the same as for the confined electron layers in the GaAs heterostructures and graphene sheets used to investigate the quantum Hall effect \cite{Janssen2012gaas}, our fermions and bosons are both relativistic, unlike in the traditional treatment \cite{zhang1989eft}. Another difference from the usual QH system is that we do not require a magnetic field. Instead, we observe a Hall conductance arising from non-equilibrium effects, where the reservoir is coupled to the system using Lindblad jump operators. This coupling demotes time-reversal invariance from a strong symmetry of the system (i.e. a symmetry which commutes with both the Hamiltonian and the jump operators) to a weak symmetry (i.e. a symmetry which commutes only with the whole Lindbladian) \cite{buca_2012}. Although the action density still possesses TRI, the interactions between the fermions, the bosons and the reservoir nonetheless lead to a TR-breaking self-energy for the fermions. When the system is probed by an external field, the self-energy gives rise to a Chern-Simons term. This in turn leads to a non-zero Hall conductivity when considering the imaginary part of the self-energy. The overall structure of our computation is the following:
\begin{enumerate}
    \item We introduce our system, consisting of fermions and bosons in a 2D material. The particle types are coupled using Lindblad operators describing dissipative effects induced by an external reservoir, in addition to non-dissipative interaction terms. 
    \item Using the above, we compute the self-energy and the dressed fermion propagators. 
    \item The dressed propagators are used to find the polarization tensor. This gives rise to a Chern-Simons term and a Hall conductance.
\end{enumerate}

We now proceed to briefly review the Keldysh formalism in section \ref{Keldysh}, followed by a more detailed description of our model in section \ref{sec:action}. This includes the non-perturbed fermionic subsystem as well as the coupling to the bosonic environment, where the latter is seen to follow from a Lindbladian. The Keldysh machinery is then employed to derive the components of the (matrix-valued) self-energy in section \ref{sec:self}. We also compute the components of the dressed propagator, which follow from the self-energy. Using this dressed propagator, we demonstrate in section \ref{sec:pol} that the polarization tensor contains a Chern-Simons term. We interpret our findings in terms of Hall physics, anomalies, and breakdown of topological invariants in the discussion, section \ref{sec:disc}. Throughout the paper, key concepts are reviewed in the appropriate sections. Various calculations are shown in more detail in the Appendices.

\section{Theoretical approach}
We first give a brief review of the basics of the Schwinger-Keldysh formalism in section \ref{Keldysh}. We subsequently describe our setup using this formalism in section \ref{sec:action}, where we state the various terms in the action and discuss the Lindblad jump operators encoding the dissipative effects. 
\subsection{Review of the Schwinger-Keldysh formalism}\label{Keldysh}

Because our description involves dissipation, Lindblad-type interactions, and a system which need not be close to equilibrium, the equilibrium quantum field theory framework of the Matsubara formalism \cite{matsubara1955new} is insufficient. Instead, we employ the Schwinger-Keldysh formalism, which holds great utility for non-equilibrium physics \cite{schwinger1960special,keldysh1964diagram,kadanoff1962quantum}. It features a closed time contour with a forward and a backward branch. The fields of the theory are denoted with a sign showing whether they live on the forward branch (e.g. $\phi^+$) or the backward branch ($\phi^-$). 
On the Keldysh contour we take time reversal to act as 
\begin{equation}\label{timerev}
    \mathbb{T}\begin{bmatrix}
        \mathcal{O}_+(t)\\ \mathcal{O}_-(t)
    \end{bmatrix}=\begin{bmatrix}
        \mathcal{T}\mathcal{O}_-(t)\\\mathcal{T}\mathcal{O}_+(t)
    \end{bmatrix},
\end{equation}
where $\mathcal{T}$ is the usual non-Keldysh time-reversal operator. It maps $t\rightarrow -t$ but otherwise acts trivially on a real scalar. It acts as $i\sigma_yK$, $K$ denoting complex conjugation, on Dirac fermions in our representation of the gamma matrices, in addition to changing the sign of $t$. Note that the two contours are exchanged. Other versions of time reversal on the Keldysh contour exist in the literature: see e.g. Refs. \cite{sieberer_keldysh_2016,haehl_schwinger-keldysh_2017}. We remark that the Keldysh action of a time-reversal invariant system receives a minus sign under time-reversal, so that $iS\rightarrow iS$.

A priori, the closed contour would give four types of two-point functions since either of the two fields could be on either half of the contour; however, these are linearly dependent, and one can work with linear combinations of the fields to reduce the number of Green's function components to three. This change of variables is known as the \textit{Keldysh rotation}. The Keldysh rotation used in this work is
\begin{eqnarray}\label{rotate}
    \phi_{\pm}=\frac{\phi_{cl}\pm \phi_q}{\sqrt{2}},\quad \psi_{\pm}=\frac{\psi_{cl}\pm \psi_q}{\sqrt{2}},
\end{eqnarray}
for bosonic fields $\phi$ and fermionic fields $\psi$. The indices $cl$ and $q$ stand for \textit{classical} and \textit{quantum}\footnote{Note that the classical-quantum nomenclature is used in the Keldysh formalism for historical reasons, and is not related to dynamical and background fields.}. The fermionic fields are sometimes instead given subscripts $1,2$, as Grassmann valued variables cannot be considered classical \cite{kamenev}. In the current work, we instead use the subscripts $1,2$ to indicate spatial coordinates.

Although not the default convention, the choice to use the bosonic Keldysh rotation also for the fermionic fields is entirely consistent (although one must not use the fermionic convention for the bosonic fields \cite{kamenev}), and results in the fermionic Green's function $G$ and its bosonic counterpart $D$ having the same structure, i.e.  
\begin{eqnarray}\label{greens}
    G=\begin{bmatrix}
        G^K& G^R\\
        G^{A}&0
    \end{bmatrix},\quad  D=\begin{bmatrix}
        D^K& D^R\\
        D^{A}&0
    \end{bmatrix}.
\end{eqnarray}
Here, the three remaining Green's function components
have subscripts $K,R,A$ for \textit{Keldysh}, \textit{retarded} and \textit{advanced}, respectively. These components obey several useful identities, including the relations
\begin{eqnarray}\label{propids}
    (G^R)^{\dagger}=G^{A}\Leftrightarrow (G^A)^{\dagger}=G^{R},\qquad (G^K)^{\dagger}=-G^{K}.
\end{eqnarray}
The matrix-valued Green's function enters into the Keldysh action as follows: an action quadratic in the fields can be written
\begin{eqnarray}\label{Sgeneric}
    S=\int d^2xdt\;\begin{bmatrix}
        \psi^{\dagger}_{cl}&\psi^{\dagger}_q
    \end{bmatrix}
    G^{-1}\begin{bmatrix}
        \psi_{cl}\\
        \psi_q
    \end{bmatrix},
\end{eqnarray}
for a fermionic system. Bosons are treated analogously, with the replacement $G^{-1}\rightarrow D^{-1}$. 

For a more thorough introduction to the Keldysh formalism, the uninitiated reader is recommended to read e.g. \cite{kamenev}. We now use this formalism to describe our system. 

\subsection{The Keldysh action}\label{sec:action}
We now describe our unperturbed system, and outline how it is affected by dissipation using a Lindbladian description. Throughout, we use the Minkowski gamma matrices 
\begin{equation}\label{minkowski}
    \gamma_0=\sigma_z, \quad \gamma_1=i \sigma_y, \quad  \gamma_2=-i \sigma_x,
\end{equation}
such that the Clifford algebra $\left\{\gamma_\mu, \gamma_\nu\right\}=2\eta_{\mu, \nu}$, with signature $(1, -1, -1)$ is satisfied, as well as the identity
\begin{eqnarray}\label{epsilonid}
    \mathrm{Tr}(\gamma_{\mu}\gamma_{\nu}\gamma_{\rho})=-2i\epsilon_{\mu\nu\rho}.
\end{eqnarray}
In addition, we set $\hbar=c=1$.

Our first task is to construct a consistent action for 2d gapped fermions where time-reversal symmetry is not broken. To accomplish this, we include a relativistic Weyl fermion $\Psi=(\psi_{\uparrow},\psi_{\downarrow})^T$, consisting of two chiral fermions $\psi_{\uparrow},\psi_{\downarrow}$, and a relativistic real boson $\phi$.

A consistent action which can accommodate dissipation may be constructed using the Lindblad equation \cite{BreuerPetruccione2002},
\begin{equation}\label{lindbladeq}
    \frac{d\rho}{dt}=-i[H,\rho]+\sum_j\int d^2x\ l_j(x)\rho l^{\dagger}_j(x)-\{l^{\dagger}_j(x)l_{j}(x),\rho\}.
\end{equation}
This equation generates trace-preserving, completely positive Markovian time evolution.
The jump operators $l_j$ are so far generic, and will be specified later. The trace of the density matrix $\rho$ equals the partition function
\begin{equation}\label{parttn}
    Z=\int \mathcal{D}[\psi_+^{\dagger},\psi_-^{\dagger},\psi_+,\psi_-,\phi_+,\phi_-]e^{i\mathcal{S}},
\end{equation}
where the subscript indicates whether the field lives on the forward ($+$) or backward ($-$) branch.

As shown in Refs.~\cite{sieberer_keldysh_2016,sieberer_universality_2025}, the Keldysh action $\mathcal{S}$ is given by

\begin{equation}\label{laction2}
    \mathcal{S}=\int d^2xdt\  \psi^{\dagger}_+i\partial_t \psi_+-\psi^{\dagger}_- i\partial_t \psi_-+(\partial_t\phi_+)^2-(\partial_t\phi_-)^2-i\mathcal{L}.
\end{equation}
The Lindbladian in the last term is given by 
\begin{align}\label{Lterm}
    \mathcal{L}&=-i(\mathcal{H}_+-\mathcal{H}_-) \nonumber\\&+\sum_j\bigg[\ell_{j+}\ell_{j-}^{\dagger}-\frac{1}{2}\left(\ell^{\dagger}_{j+}\ell_{j+}+\ell^{\dagger}_{j-}\ell_{j-}\right)\bigg].
\end{align}
Here, the Hamiltonian dynamics is included in the Hamiltonian density $\mathcal{H}_{\pm}$ while the dissipation is encoded in the $\ell_{j\pm}$. The latter are the fields in the partition function corresponding to the $l_j$ operators in eq. \eqref{lindbladeq}. Now, the action may be split into three parts:
\begin{equation}
    \mathcal{S}=\mathcal{S}_0+\mathcal{S}_{int}+\mathcal{S}_{diss}
\end{equation}
where $\mathcal{S}_0$ is the non-dissipative quadratic part of the action, $\mathcal{S}_{int}$ includes all non-dissipative interactions, and $\mathcal{S}_{diss}$ contains all dissipation. $\mathcal{S}_0$ may be further split into a fermionic and a bosonic part, the fermionic part being given by
\begin{equation}\label{S_F}
 \mathcal{S}_{F,0}=i\int d^2xdt \begin{bmatrix}
        \psi_{cl}^{\dagger} &\psi^{\dagger}_q
    \end{bmatrix}
    \begin{bmatrix}
        0 &\partial -i\epsilon\\
        \partial+i\epsilon & \epsilon F_F
    \end{bmatrix}
    \begin{bmatrix}
        \psi_{cl} \\\psi_q,
    \end{bmatrix}
\end{equation}

where $\partial=\sigma^\mu\partial_\mu$\footnote{This should not be mistaken for the holomorphic derivative.}, as well as $\sigma^{\mu}=(I, \sigma_x, \sigma_y)$, and $\partial_{\mu}=(\partial_t, -v_F\partial_x, -v_F\partial_y)$, 
after a Keldysh rotation. The parameter $\epsilon$ is an infinitesimal regularization, and $F_F$ encodes the initial state of the unperturbed fermions, which will not play a role due to the Markovian nature of the full action. Note that the unperturbed fermions are massless.

The bosonic part of $\mathcal{S}_0$ is given by 
\begin{widetext}
\begin{align}\label{SB0}
    \mathcal{S}_{B,0}&=\frac{1}{2}\int d^2xdt \begin{bmatrix}
        \phi_{cl} &\phi_q
    \end{bmatrix}
    \begin{bmatrix}
        0 & (\partial_t+i\epsilon)^2-c_s^2\nabla^2+m^2\\
       (\partial_t-i\epsilon)^2-c_s^2\nabla^2+m^2 & i\epsilon F_B
    \end{bmatrix}
    \begin{bmatrix}
        \phi_{cl}\\\phi_q
    \end{bmatrix},
\end{align}
\end{widetext}
where $F_B$ encodes the initial state of the unperturbed bosons. From the above, it follows that we may write the bosonic dispersion relation as
\begin{eqnarray}\label{dispbos}
    \omega(k)=\pm \sqrt{c_s^2k^2+m^2}
\end{eqnarray}
in the limit where the regularization $\epsilon\rightarrow 0$.
The non-dissipative interacting part of the action is given by
\begin{align}\label{S_H}
    \mathcal{S}_{int}=g_S\int d^2xdt(&\psi^{\dagger}_+\psi_+\phi_+-\psi^{\dagger}_-\psi_-\phi_-)\nonumber\\=\frac{1}{\sqrt{2}}g_S\int d^2xdt\big(&\psi^{\dagger}_{cl}\psi_{cl}\phi_{q}+\psi^{\dagger}_{cl}\psi_{q}\phi_{cl}\nonumber\\+&\psi^{\dagger}_{q}\psi_{cl}\phi_{cl}+\psi^{\dagger}_{q}\psi_{q}\phi_{q}\big),
\end{align}
which is a density-density interaction mediated by the scalar field $\phi$. Finally, the jump operators are taken to add or remove fermions in a chirality-dependent way. Letting the operator index $j=\alpha=\pm 1$ label the two components of the Weyl fermion, with $\alpha=1$ ($-1$) for the $\uparrow$ ($\downarrow$) component, the jump operators are
\begin{align}\label{jump}
    \begin{cases}\ell_{\alpha}&=\sqrt{g_{L}}\left(i(\sigma_z\psi)_{\alpha}+g_B\phi\psi_{\alpha}\right),\\
    \ell'_{\alpha}&=\sqrt{g_{G}}\left(-i(\sigma_z\psi^{\dagger})_{\alpha}+g_B\phi\psi^{\dagger}_{\alpha}\right).
    \end{cases}
\end{align}
The action of $\ell_{\alpha}$ ($\ell'_{\alpha}$) is to remove (add) a fermion, modulated by the scalar field. Inserting them into the Lindbladian results in several contributions to $S_{diss}$, as described in detail in Appendix \ref{sec:LindbladApp}. Only a few of these are of interest. The crucial term is the following Yukawa-like contribution,
\begin{equation}\label{SD}
    S_{D}=ig_B\int d^2xdt\  (\phi_--\phi_+)(g_L\bar{\psi}_-\psi_++g_G\bar{\psi}_+\psi_-),
\end{equation}
which together with $S_{int}$ produces a Dirac mass for the fermions, as we will see. Here the fermionic “bar fields" $\bar{\psi}_{cl,q}=\psi^{\dagger}_{cl,q}\gamma_0=\psi^{\dagger}_{cl,q}\sigma_z$ for both components of $\Psi$; c.f. \eqref{minkowski}.  Note that $S_D$ \textit{does not} violate time-reversal invariance. This is due to the extra $i$ in the path integral \eqref{parttn}, as well as $g_G$ and $g_L$ being exchanged under time reversal since gain and loss are exchanged. Let us define the coefficients
\begin{eqnarray}\label{gpm}
    g_{\pm}=\frac{g_G\pm g_L}{2}
\end{eqnarray}
as the average and half-difference of the gain and loss coefficients. We note that $g_+$ and $g_-$ are even and odd under time reversal, respectively. Using the coefficients \eqref{gpm}, we may rewrite \eqref{SD} in terms of the Keldysh rotated fields in \eqref{rotate}, giving 
\begin{align}\label{SB'}
    \mathcal{S}_D=-\sqrt{2}ig_B\int d^2xdt\ g_+&(\psi^{\dagger}_{cl}\sigma_z\psi_{cl}-\psi_q^{\dagger}\sigma_z\psi_q)\phi_q\nonumber\\+g_-&(\psi^{\dagger}_q\sigma_z\psi_{cl}-\psi^{\dagger}_{cl}\sigma_z\psi_q)\phi_q.
\end{align}
This is shown in more detail in Appendix \ref{sec:LindbladApp}. Furthermore, the dissipation modifies the quadratic part of the action for the fermions, yielding (c.f. Appendix \ref{sec:LindbladApp} for a derivation)
\begin{equation}\label{eq:G0}
     \mathcal{S}_F=i\int d^2xdt \begin{bmatrix}
        \psi_{cl}^\dagger &\psi_q^\dagger
    \end{bmatrix}
    \begin{bmatrix}
        0 &\partial-ig_+\\
       \partial+ig_+ & -2ig_-
    \end{bmatrix}
    \begin{bmatrix}
        \psi_{cl}\\
        \psi_q
    \end{bmatrix}.
\end{equation}
By comparison with \eqref{Sgeneric}, the quadratic form above is the inverse of the bare propagator $G_0$, so inverting lets us state for later use that 
\begin{align}\label{G0}
\hspace{-0.3
cm} G_0&=\begin{bmatrix}
     G^K_0 & G^R_0\\
     G^{A}_0&0
    \end{bmatrix}\nonumber\\
\hspace{-0.3cm}    &=\frac{1}{(\omega-H)^2+g_+^2}\begin{bmatrix}
        2ig_-& \omega-H-ig_+\\
        \omega-H+ig_+&0
    \end{bmatrix}.
\end{align}


We remark that the action as a whole preserves TR invariance. This follows from the Keldysh rotation \eqref{rotate}: the quantum field $\phi_q=\frac{\phi_+-\phi_-}{\sqrt{2}}$ transforms as $\phi_q\rightarrow -\mathcal{T}\phi_q=-\phi_q$ under time reversal, from eq. \eqref{timerev}. Similarly, $\psi_q$ picks up a minus sign and a factor of $i\sigma_y$ which cancels against the factor of $-i\sigma_y$ from the adjoint field. The end result is that all terms with an odd number of quantum fields pick up an overall minus sign, which cancels the overall minus sign from the rest of the action. Because most of our interaction terms have an odd number of quantum fields, 
they respect TRI. The exception is the second line of \eqref{SB'}, which has an even number of quantum fields, but is still TR invariant because gain $g_G$ and loss $g_L$ are exchanged under TRI, so that $g_-\rightarrow -g_-$. The non-interacting terms \eqref{eq:G0} and \eqref{SB0} are seen to also obey TRI, due to the factors of $i$ multiplying the $q-q$ components.

\section{The self-energy}\label{sec:self}
As seen in the previous section, the fermions in our system interact with a bosonic field and with a reservoir. Due to the interactions, the free propagator will be modified by various increasingly complicated scattering processes \cite{coleman}. In particular, these processes can generate a mass for the propagating particle. Adding all these contributions still gives a closed-form expression, and the \textit{dressed} propagator $\mathcal{G}$ (where interactions are included) and the \textit{bare} propagator $\mathcal{G}_0$ (where they are not) are related by the Dyson equation: $\mathcal{G}=\mathcal{G}_0+\mathcal{G}_0\mathit{\Sigma}\mathcal{G}$. Here $\mathit{\Sigma}$ is the self-energy, which contains all one-particle irreducible processes, and acts as a mass term in $\mathcal{G}$. Hence it gives a modified mass for the particles with bare propagator $\mathcal{G}_0$.  

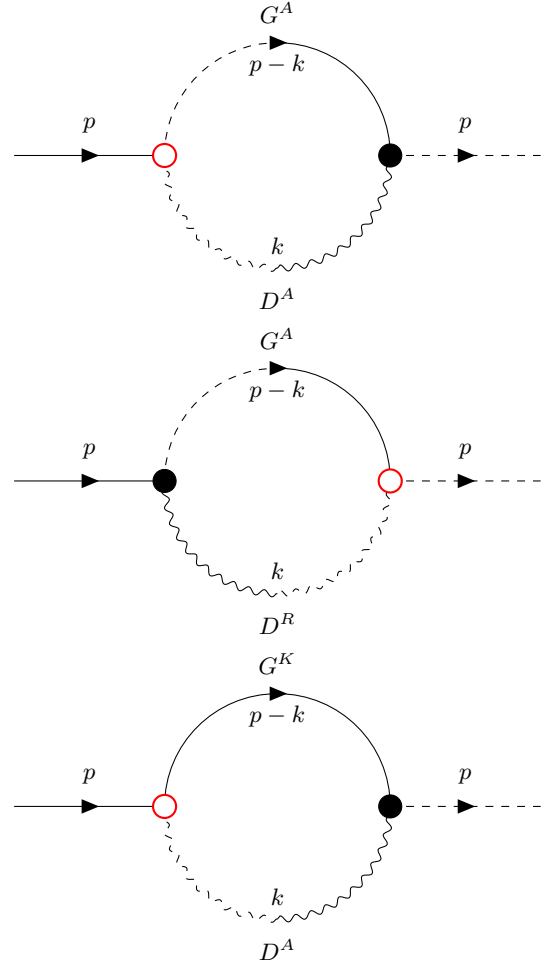
\begin{figure}
    \centering

\begin{tikzpicture}
\begin{feynman}
  \vertex (a) at (0,2);
  \vertex (x) at (2,2);
  \vertex (y) at (5,2);
  \vertex (mid) at (3.5,3.5);
  \vertex (z) at (3.5,0.5);
  \vertex (b) at (7,2);

  \diagram*{
    (a) -- [plain,with arrow=0.5] (x),
    (x) -- [plain,dashed,bend left=45] (mid),
    (mid) -- [plain,with arrow=0,label=above:\(x\),bend left=45] (y),
    (y) -- [plain, dashed,with arrow=0.5] (b),
     (y) -- [boson ,bend left=45] (z),
    (z) -- [boson,dashed,bend left=45] (x)
  };
   
\node at ({1,2.4}) {$p$};
\node at ({6,2.4}) {$p$};
\node at ({3.5,3.2}) {$p-k$};
\node at ({3.5,3.9}) {$G^{A}$};
\node at ({3.5,0.8}) {$k$};
\node at ({3.5,0.1}) {$D^{A}$};
\node [circle,red,draw=red,thick,fill=white] (e) at (2,2);
\node [circle,black,draw=black,fill=black] (f) at (5,2);


\end{feynman}
\end{tikzpicture}
\begin{tikzpicture}
\begin{feynman}
  \vertex (a) at (0,2);
  \vertex (x) at (2,2);
  \vertex (y) at (5,2);
  \vertex (mid) at (3.5,3.5);
  \vertex (z) at (3.5,0.5);
  \vertex (b) at (7,2);

  \diagram*{
    (a) -- [plain,with arrow=0.5] (x),
    (x) -- [plain,dashed,bend left=45] (mid),
    (mid) -- [plain, with arrow=0,label=above:\(x\),bend left=45] (y),
    (y) -- [plain, dashed,with arrow=0.5] (b),
     (y) -- [boson,dashed,bend left=45] (z),
    (z) -- [boson,bend left=45] (x)
  };

\node at ({1,2.4}) {$p$};
\node at ({6,2.4}) {$p$};
\node at ({3.5,3.2}) {$p-k$};
\node at ({3.5,3.9}) {$G^{A}$};
\node at ({3.5,0.8}) {$k$};
\node at ({3.5,0.1}) {$D^{R}$};
\node [circle,black,draw=black,fill=black] (e) at (2,2);
\node [circle,red,draw=red,thick,fill=white] (f) at (5,2);


\end{feynman}
\end{tikzpicture}

\begin{tikzpicture}
\begin{feynman}
  \vertex (a) at (0,2);
  \vertex (x) at (2,2);
  \vertex (y) at (5,2);
  \vertex (mid) at (3.5,3.5);
  \vertex (z) at (3.5,0.5);
  \vertex (b) at (7,2);

  \diagram*{
    (a) -- [plain,with arrow=0.5] (x),
    (x) -- [plain,bend left=45] (mid),
    (mid) -- [plain, with arrow=0,label=above:\(x\),bend left=45] (y),
    (y) -- [plain, dashed,with arrow=0.5] (b)
  };
  \diagram*{
    (y) -- [boson ,bend left=45] (z),
    (z) -- [boson,dashed,bend left=45] (x)
  };
\node at ({1,2.4}) {$p$};
\node at ({6,2.4}) {$p$};
\node at ({3.5,3.9}) {$G^K$};
\node at ({3.5,3.2}) {$p-k$};
\node at ({3.5,0.8}) {$k$};
\node at ({3.5,0.1}) {$D^{A}$};
\node [circle,red,draw=red,thick,fill=white] (e) at (2,2);
\node [circle,black,draw=black,fill=black] (f) at (5,2);


\end{feynman}
\end{tikzpicture}

\caption{The diagrams that give chirality-mixing contributions to the self-energy component $\Sigma^{A}$, to one-loop order. The external legs have been included to clarify which propagator type is ingoing or outgoing, even though we only compute the amputated diagram. We use a convention with dashed (solid) lines for quantum (classical) fields to more clearly show the propagators that we are using.}
\label{fig:SigmaA}
\end{figure}

In the Keldysh formalism, the presence of indices $cl,q$ gives the self-energy a $2\times 2$ matrix structure. The sum over interactions leads to a modified Dyson equation on the form
\begin{eqnarray}\label{dyson}
    G=G_0+G_0\circ \Sigma\circ G,
\end{eqnarray}
where $\circ$ indicates a convolution of the spacetime coordinates in addition to matrix multiplication\footnote{We remind the reader that the dressed and bare propagators both have a matrix structure.}.
In particular, since we use the Keldysh rotation \eqref{rotate}, it follows that the self-energy $\Sigma$ has the structure \cite{kamenev}
\begin{eqnarray}
    \Sigma=\begin{bmatrix}
        0 & \Sigma^{A}\\
        \Sigma^R & \Sigma^K
    \end{bmatrix}.
\end{eqnarray}
We now compute this matrix valued self-energy. Since it is a renormalization contribution to the electron mass, we need to consider which one-loop diagrams depend on chirality, i.e. which diagrams have an overall matrix factor of $\sigma_z$. By considering our interaction vertices eqs. \eqref{S_H} and \eqref{SB'}, we arrive at the one-loop Feynman diagrams in Fig. \ref{fig:SigmaA} 
for the advanced component $\Sigma^{A}$. After a long but straightforward calculation -- which is outlined in more detail in Appendix \ref{sec:selfenergyApp} -- the result (to first order in $\nu$) is that $\Sigma^{A}=(\Sigma_r-i\Sigma_i)\sigma_z$ with
\begin{equation}\label{sigmar}
    \Sigma_r=\frac{g_Sg_Bg_-}{4\pi}\bigg[\frac{2\Lambda^2}{g_+^2+v_F^2\Lambda^2}-\frac{1}{v_F^2}\ln\bigg(1+\frac{v_F^2\Lambda^2}{g_+^2}\bigg)\bigg]
\end{equation}
and
\begin{eqnarray}\label{sigmai}
    \Sigma_i=\frac{g_Sg_Bg_-}{2\pi g_+}\frac{\Lambda^2(g_+^2-v_F^2\Lambda^2)}{(g_+^2+v_F^2\Lambda^2)^2}\nu,
\end{eqnarray}
with $\Lambda$ a UV cutoff for the boson momentum. A few further assumptions have been made here: firstly, an expansion has been made to first order in the frequency $\nu$, and the external momentum has been set to $0$. This corresponds to an electron gas held at low temperature. Secondly, the bosons have been assumed to be extremely light or massless ($m=0$ in \eqref{dispbos}) and the boson velocity parameter $c_s$ has been assumed to be small compared to the Fermi velocity $v_F$. 

Since the Feynman diagrams for the retarded component $\Sigma^R$ are manifestly the Hermitian adjoints of the ones for the advanced component -- see Fig. \ref{fig:SigmaR} in Appendix \ref{sec:selfenergyApp} --  it immediately follows that 
$    \Sigma^R=(\Sigma^{A})^{\dagger}=(\Sigma_r+i\Sigma_i)\sigma_z,
$
with $\Sigma_r,\Sigma_i$ given by eqs. \eqref{sigmar} and \eqref{sigmai}. The Keldysh component instead has four diagrams (c.f. Fig. \ref{fig:SigmaK}). We note that their sum is manifestly anti-Hermitian, meaning that $\Sigma^K$ is as well. An arduous but straightforward computation of the diagrams (see Appendix \eqref{sec:selfenergyApp} for details) eventually gives the Keldysh component 
\begin{align}\label{SigmaK}
    \Sigma^K&=\frac{1}{2\pi}g_Sg_B\sigma_z\bigg[1+\frac{g_-^2}{g^2_+}
    \bigg]\frac{i\Lambda^2(g_+^2-v_F^2\Lambda^2)}{(g_+^2+v_F^2\Lambda^2)^2}\nu\nonumber\\
    &=\bigg[\frac{g_+}{g_-}+\frac{g_-}{g_+}
    \bigg]\sigma_zi\Sigma_i\equiv \chi\times \sigma_zi\Sigma_i,
\end{align}
which is indeed anti-Hermitian. Here, the coefficient $\chi$ has been defined for later use. 

From the Dyson equation \eqref{dyson} and matrix inversion of the bare propagator $G_0$, it follows that the dressed propagator obeys
\begin{align}\label{Ginvert}
    G^{-1}&=G_0^{-1}-\Sigma\nonumber\\
    &=\begin{bmatrix}
        0 & [G_0^{A}]^{-1}-\Sigma^{A}\\
        [G_0^R]^{-1}-\Sigma^{R}& -[G_0^{R}]^{-1}G_0^K[G_0^{A}]^{-1}-\Sigma^K
    \end{bmatrix}
    \nonumber\\
        &=\begin{bmatrix}
        0 & [G_0^{A}]^{-1}-\Sigma^{A}\\
        [G_0^R]^{-1}-\Sigma^{R}& -\big(2ig_-+\Sigma^K\big)
    \end{bmatrix},
\end{align}
where the last equality follows from  \eqref{G0}.
By inverting the above and comparing to the structure \eqref{greens}, one may identify the components of the dressed Green's function. Using the bare propagator \eqref{G0}, the retarded and advanced components simply become
\begin{align}\label{GRA}
    G^{R/A}
    &=\bigg(\big(G_0^{R/A}\big)^{-1}-\Sigma^{R/A}\bigg)^{-1}\nonumber\\&=\frac{1}{\omega^2-g_+^2-\Sigma_r^2+\Sigma_i^2-v_F^2k^2\pm 2i(g_+\omega-\Sigma_r\Sigma_i)}\nonumber\\
    &\times\bigg[(\omega\pm ig_+)I+(\Sigma_r\pm i\Sigma_i)\sigma_z+v_F\vec{k}\cdot \vec{\sigma}\bigg].
\end{align}
 We note that this fulfills the relations in \eqref{propids}. The Keldysh component instead becomes
 \begin{align}
     G^K&=\bigg(\big(G_0^{R}\big)^{-1}-\Sigma^{R}\bigg)^{-1}\big(2ig_-+\Sigma^K\big)\bigg(\big(G_0^{A}\big)^{-1}-\Sigma^{A}\bigg)^{-1}\nonumber\\
     &=G^R\big(2ig_-+i\chi \Sigma_i \sigma_z\big)G^{A},
 \end{align}
using eq. \eqref{GRA}. Since $\chi$ and $\Sigma_i$ are real numbers, $G^K$ is manifestly anti-Hermitian, as expected from eq. \eqref{propids}.

 By computing the matrix multiplications and expressing the result of the calculation in the Pauli basis $\{I,\sigma_x,\sigma_y,\sigma_z\}$, we find general expressions for the dressed propagators. These are rather complicated, and are stated in Appendix \ref{sec:poltensApp}, where we use them to compute the polarization tensor.

The propagators become much simpler in the limit where the external frequency $\nu=0$. This makes the self-energy component $\Sigma_i=0$ and also $\Sigma^K=0$ (see eqs. \eqref{sigmai} and \eqref{SigmaK}). 
As computed in Appendix \ref{sec:pol2app} and discussed next, this leads to a completely different result for the polarization tensor.
\section{The polarization tensor}\label{sec:pol}
To probe the system, we apply an electromagnetic field. On the Keldysh contour, this can be described by the following coupling between the fermions and the EM field: 
\begin{eqnarray}
    S_A=-\int d^2xdt \;\big( \psi^{\dagger}_+\gamma_0\slashed{A}_+\psi_+-\psi^{\dagger}_-\gamma_0\slashed{A}_-\psi_-\big),
\end{eqnarray}
where we use the representation \eqref{minkowski} of the gamma matrices. Using the Keldysh rotation \eqref{rotate}, the above is
\begin{align}
    S_{A}&=-\int d^2xdt\;\begin{bmatrix}
        \psi^{\dagger}_{cl}\gamma_0&\psi^{\dagger}_q\gamma_0\end{bmatrix}\begin{bmatrix}
            \slashed{A}_q & \slashed{A}_{cl}\\
            \slashed{A}_{cl} & \slashed{A}_{q}
        \end{bmatrix}
    \begin{bmatrix}
        \psi^{cl}\\
        \psi^q
    \end{bmatrix},
\end{align}
Writing $Z=e^{iS_A}$, one can compute
\begin{eqnarray}
    \Pi^R=\Pi^{q,cl}_{\mu\nu}=-\frac{i}{2}\frac{\delta^2Z}{\delta A_{cl}^{\mu}\delta A_{q}^{\nu}}.
\end{eqnarray}
This gives an expression for the retarded component of the polarization tensor \cite{kohri2002polarization}, which is the component related to the conductivity tensor \cite{gorbar2002magnetic,frassdorf2018abelian}:
\begin{eqnarray}\label{poltens}
     i\Pi^{q,cl}_{\mu\nu}=\frac{e^2}{2}\mathrm{Tr}\big(\gamma_{\mu}G^{R}\gamma_{\nu}G^K+\gamma_{\mu}G^K\gamma_{\nu}G^{A}\big).
     \label{eq:pol_Keldysh}
\end{eqnarray}

We wish to compute the polarization tensor \eqref{poltens} for the spatial components of $\mu,\nu$, since those components of $\Pi^{q,cl}_{\mu\nu}$ can give rise to a Chern-Simons (CS) term from the Lagrangian
\begin{eqnarray}\label{EMcoupl}
   \frac{1}{2} \int \frac{d\nu d^2p}{(2\pi)^3}A_{\mu}(p)\Pi^{q,cl}_{\mu\nu}(\nu,p)A_{\nu}(-p)+\mathcal{O}(A^3),
\end{eqnarray}
which in turn gives Hall physics. It may seem puzzling that a CS term, which is widely known to break TR invariance, arises in a system where the Lagrangian is TR invariant. However, the mass term caused by the boson-fermion coupling breaks TR symmetry when restricted to the fermion subsystem only, which is what allows the CS term to emerge despite the TRI of the action as a whole. This point is more thoroughly analyzed in the discussion below.

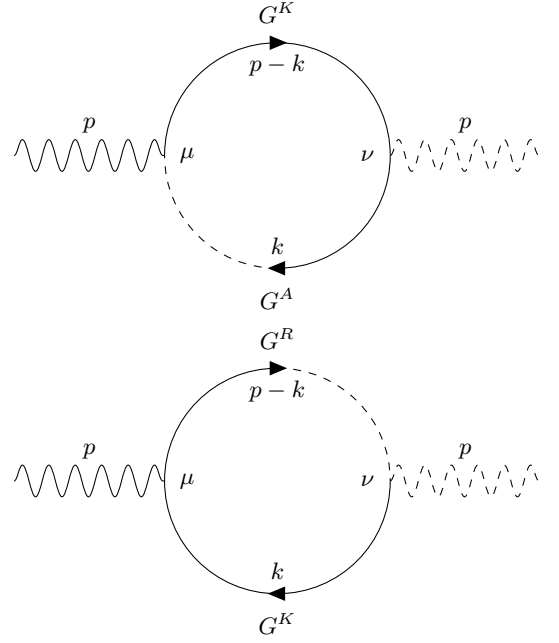
\begin{figure}
    \centering
\begin{tikzpicture}
\begin{feynman}
  \vertex (a) at (0,2);
  \vertex (x) at (2,2);
  \vertex (y) at (5,2);
  \vertex (mid) at (3.5,3.5);
  \vertex (z) at (3.5,0.5);
  \vertex (b) at (7,2);

  \draw[
  decorate,
  decoration={
    snake,
    amplitude=6pt,
    segment length=10pt}] (a) -- (x);
  \diagram*{
    (x) -- [plain,bend left=45] (mid),
    (mid) -- [plain, with arrow=0,label=above:\(x\),bend left=45] (y)
  };
  \draw[dashed,
  decorate,
  decoration={
    snake,
    amplitude=6pt,
    segment length=10pt
    }] (y) -- (b);
  \diagram*{
    (y) -- [plain ,bend left=45] (z),
    (z) -- [plain,dashed,with arrow=0,bend left=45] (x)
  };
\node at ({1,2.4}) {$p$};
\node at ({6,2.4}) {$p$};
\node at ({3.5,3.9}) {$G^K$};
\node at ({3.5,3.2}) {$p-k$};
\node at ({3.5,0.8}) {$k$};
\node at ({3.5,0.1}) {$G^{A}$};
\node (e) at (2.3,2) {$\mu$};
\node (f) at (4.7,2) {$\nu$};


\end{feynman}
\end{tikzpicture}

\begin{tikzpicture}
\begin{feynman}
  \vertex (a) at (0,2);
  \vertex (x) at (2,2);
  \vertex (y) at (5,2);
  \vertex (mid) at (3.5,3.5);
  \vertex (z) at (3.5,0.5);
  \vertex (b) at (7,2);

   \draw[decorate,
  decoration={
    snake,
    amplitude=6pt,
    segment length=10pt}] (a) -- (x);
  \diagram*{ 
    (x) -- [plain,bend left=45] (mid),
    (mid) -- [plain,dashed, with arrow=0,label=above:\(x\),bend left=45] (y),
  };
  \draw[dashed,
  decorate,
  decoration={
    snake,
    amplitude=6pt,
    segment length=10pt}] (y) -- (b);
  \diagram*{
    (y) -- [plain,bend left=45] (z),
    (z) -- [plain,with arrow=0,bend left=45] (x)
  };
\node at ({1,2.4}) {$p$};
\node at ({6,2.4}) {$p$};
\node at ({3.5,3.2}) {$p-k$};
\node at ({3.5,3.9}) {$G^{R}$};
\node at ({3.5,0.8}) {$k$};
\node at ({3.5,0.1}) {$G^{K}$};
\node [] (e) at (2.3,2) {$\mu$};
\node [] (f) at (4.7,2) {$\nu$};


\end{feynman}
\end{tikzpicture}
\caption{Feynman diagrams for the polarization tensor component $\Pi^R_{\mu\nu}$, to one-loop order.}
\label{fig:pol}
\end{figure}

Let us now compute the polarization tensor in \eqref{EMcoupl}, using the polarization tensor diagrams given in Fig. \ref{fig:pol}. We first outline the computation using the general dressed propagators with $G^{R/A}$ as in \eqref{GRA} and $G^K$ specified by \eqref{GKtot}, together with \eqref{GKchi} and \eqref{GK-}. We then outline how the computation simplifies for vanishing external frequency $\nu=0$ of the self-energy, which gives the special cases \eqref{GK2}, \eqref{GRA2} of the dressed propagators. In this limit, the self-energies are
\begin{eqnarray}
    \Sigma^R=\Sigma^{A}=\Sigma_r\sigma_z,\quad \Sigma^K=0,
\end{eqnarray}
to one-loop order. Therefore, this limit corresponds to keeping the mass terms (which do not depend on $\nu$) but not the wavefunction renormalization (which does).
It is seen that in this special case, the components of interest vanish.

For the case where the frequency-dependent terms in the self-energies are kept, the derivation is shown in greater detail in Appendix \ref{sec:poltensApp}. Here, we only outline a few of the key steps. From the trace in eq. \eqref{poltens} and the factors $\gamma_{\mu},\gamma_{\nu}$ with $\mu,\nu\in\{1,2\}$, it follows that the only terms in the propagators contributing to the result are those giving an overall factor $\sigma_z$. In other words, for each diagram, the nonzero contributions are those where one of the propagators gives a factor $I$ and the other gives a factor $\sigma_z$. Secondly, we find that the two diagrams are minus each other's complex conjugates, meaning that the result of one follows from the result of the other.
Expanding our integrand to first order in the external momentum components $\vec{p}=(p_0,p_1,p_2)^T=(\nu,-v_F p_x,-v_Fp_y)^T$ to isolate the CS contribution, straightforward but involved calculations eventually yield  
the polarization tensor 
\begin{align}\label{pol}
    \Pi_{\mu\nu}=- ie^2p_{\rho}\epsilon^{\mu\rho\nu}\frac{g_Bg_S}{128\pi^2v_F^4}
,\qquad \mu,\nu\in\{1,2\},
\end{align}
in the limit $g_-\rightarrow 0$, i.e. in the limit where the system gains and loses fermions equally fast; c.f. eqs. \eqref{jump} and \eqref{gpm}. Here, $\rho=0,1,2$ and the frequency $\nu=p_0$. Consequently, the coupling \eqref{EMcoupl} between the fermions and the applied electric field becomes a CS term, i.e. it takes the form of the effective action
\begin{eqnarray}
    \Gamma_{CS}[A]=\frac{k}{4\pi}\epsilon^{\mu\nu\rho}\int d^2xdt\; A_{\mu}\partial_{\nu}A_{\rho}
    \label{eq_Gamma_CS}
\end{eqnarray}
once the expression for the polarization tensor \eqref{pol} is inserted together with the replacement $p_{\rho}\rightarrow -i\partial_{\rho}$. Its level is 
\begin{eqnarray}
    k=-e^2 \frac{\mathrm{g_Sg_B}}{32\pi v_F^4}.
\end{eqnarray}
From standard results on CS theory, it follows that the above describes a system with Hall conductivity 
\begin{eqnarray}\label{cond}
    \sigma_{xy}=\frac{k}{2\pi}=-e^2 \frac{\mathrm{g_Sg_B}}{64\pi^2 v_F^4}.
\end{eqnarray}
We note that the sign of the conductivity is determined by the couplings $g_B$ and $g_S$, but is indifferent to the sign of the gain-loss imbalance $g_-$.

The above computation is more concise and transparent when done using the simplified Green's functions for vanishing frequency in the self-energy (see eqs. \eqref{GRA2} and \eqref{GK2}). This eventually leads to the Chern-Simons level being given by 
\begin{widetext}
\begin{equation}\label{respol2}
   k=16\pi g_-\Sigma_r \int_{-\infty}^{\infty} \frac{d\omega}{2\pi}\int\frac{d^2q}{(2\pi)^2}\frac{\omega(g_+^2-\Sigma_r^2-v_F^2q^2+\omega^2)}{\left[\left(\omega+ i g_+\right)^2-v_F^2q^2-\Sigma_r^2\right]^2\left[\left(\omega-i g_+\right)^2-v_F^2q^2-\Sigma_r^2\right]^2}=0,
\end{equation}
\end{widetext}
where the last equality holds because the integrand is odd in $\omega$, and is integrated over an even interval. Thus, the Chern-Simons term vanishes for vanishing external frequency $\nu$ in the self-energy. This computation is shown in more detail in Appendix \ref{sec:pol2app}.


We remind the reader that in the equilibrium case, the polarization tensor for $(2+1)$-dimensional fermions becomes \cite{MulliganBurnell2013,arouca2022anomalies}
\begin{eqnarray}\label{csherm}
    \Pi_{\mu\nu}(p)=-\frac{1}{4\pi}p_{\alpha}\epsilon^{\alpha\mu\nu}\frac{M}{|M|}-(p^2\delta_{\mu\nu}-p_{\mu} p_{\nu})\frac{1}{16|p|},
\end{eqnarray}
in the limit where the fermions are massless. Here, $M$ is the mass of a spinor field added as UV regularization. Thus, the first term gives a Chern-Simons term, where the level $k_H$ is \cite{arouca2022anomalies} 
\begin{eqnarray}\label{kH}
    k_{H}=\frac{ie^2}{2}\frac{M}{|M|}=\pm \frac{ie^2}{2}, 
\end{eqnarray}
depending on the sign of the regularization. The above is half-quantized, unlike in the $m\rightarrow \infty$ limit where integer quantization occurs. The massless case is thus a form of the parity anomaly, since it violates the integer quantization of the Hall conductance for free fermions \cite{niemi_axial-anomaly-induced_1983, redlich_gauge_1984, alvarez-gaume_gravitational_1984, haldane1988aqhe, arouca2022anomalies}. 

By contrast to the equilibrium case \eqref{kH}, our result \eqref{respol2} for vanishing self-energy frequency is identically zero, regardless of whether the mass term $\Sigma_r$ is taken to zero or not. When wave-function renormalization is taken into account (i.e. $\Sigma_i\propto \nu$ is kept), we find a nonvanishing Hall conductivity \eqref{cond}, but it is not quantized -- not even in half-integers. This is especially interesting considering that from eqs. \eqref{sigmar} and \eqref{sigmai}, the self-energy vanishes in the limit\footnote{Note that in our setup, the coefficients $g_G$ and $g_L$ set the chirality, which is related to the sign of $g_-$ via \eqref{gpm}. This is similar to the role of an infinitesimal mass term in the usual context of the parity anomaly. By contrast, setting either $g_S$ or $g_B$ to zero would make the effect vanish.} $g_-\rightarrow 0$. Therefore, this limit is an analogue of the massless limit $m\rightarrow 0$ in the Hermitian case, which makes it surprising that the half-quantization of \eqref{kH} is absent.  

\section{Discussion}\label{sec:disc}
Starting with a completely time-reversal invariant dissipative action, we obtain a renormalized fermionic theory with a finite mass term (corresponding to the real part of the fermionic self-energy) after integrating out the bosonic degrees of freedom. Although the expectation is that this will lead to a finite Hall conductivity (which becomes half-quantized in the limit of infinitesimal mass \cite{niemi_axial-anomaly-induced_1983, redlich_gauge_1984,alvarez-gaume_gravitational_1984, haldane1988aqhe, arouca2022anomalies}), we found that this mass term leads to a vanishing Hall conductivity. On the other hand, considering both real and imaginary contributions to the self-energy (with the latter related to the loss of spectral weight of fermionic quasiparticles) leads to an unquantized Hall effect. In particular, it depends explicitly on the coupling constants such that it goes to zero for the free theory, and it is linearly proportional to the couplings even in limits where the self-energies (both real and imaginary parts) vanish. This is the main result of our work: that the generation of a Hall effect by a mass term in an out-of-equilibrium setting can be drastically different from the usual equilibrium case.

The parity anomaly is absent when we only consider the real part of the self-energy, corresponding to the usual setup with a Dirac fermion which has a mass term. When considering an imaginary self-energy correction, we instead have a version of the parity anomaly  with a coefficient that is unquantized. These results are due to the modification of the Keldysh component of the Green's function, which accounts for the change of density matrix of the system. We remark that a calculation using the modified retarded and advanced Green's functions with a Keldysh component determined by a Fermi-Dirac distribution would lead to the regular parity anomaly in the limit where gain and loss (and therefore the mass term) vanish. Hence, the effect that we consider can only be taken into account using the Keldysh formalism, which accounts for the change of the density matrix of the system, and would not be captured by an effective (non-Hermitian) Hamiltonian approach.

An unquantized Chern-Simons level is problematic when quantizing the electromagnetic field, since it is then not invariant under large gauge transformations. Although this breaking could be compensated by a chiral anomaly of edge states by an anomaly inflow mechanism \cite{CALLAN1985427, MulliganBurnell2013, arouca2022anomalies}, this makes the theory inconsistent when defining it on a surface without boundaries, such as a torus. Although in practice one expects these cases to be less experimentally relevant, it is still a problem in defining the theory more broadly. Nonetheless, together with an anomaly inflow mechanism, our results are unproblematic in the 2D plane, which is the geometry most closely resembling an actual QH sample.

It is also worth mentioning that the lack of quantization is not only found in non-equilibrium settings. For a massive Dirac theory at equilibrium, one obtains quantized values just for the massless limit or when the mass is big enough to induce a stable gap and obtain the IQHE \cite{MulliganBurnell2013,arouca2022anomalies}. This is what was measured in, for example, Ref.~\cite{mogi_experimental_2022} in which a magnetic field controls the magnitude of a mass term induced at the surface of a magnetic topological insulator. In our case, however, we do not obtain a quantized value in either of these limits.

The dependence of \eqref{cond} on the coupling $g_B$ is a manifestation of how chiral symmetry is broken. Since the bosons couple to the fermions in a chirality-dependent way, the bosons are expected to introduce chirality to the system. This chirality depends on the sign of $g_B$. In general, the breaking of chiral symmetry coincides with the breaking of time-reversal invariance in the quantum Hall context, and can be seen e.g. in the edge currents of a QH sample. In the model considered above, however, chirality is introduced through non-equilibrium contributions while retaining time-reversal as a weak symmetry in the bare action.


The point that the mass term in the dressed action breaks TRI despite the bare one respecting TR symmetry is a subtle one, and so we discuss it more thoroughly here. We note that the interaction term \eqref{SB'} is TR invariant because the minus signs $i$, $g_-$, $\psi_q$ and $\psi_q^{\dagger}$ pick up under $\mathbb{T}$ are balanced by the minus sign of $\phi_q$. Thus, the TRI only holds when considering the \textit{entire} system. By only considering the reservoir through the mass term it leads to, we are considering an effective description where the bosonic part of the system has been traced out. Thus, the lack of TRI applies to the fermionic subsystem but not the system as a whole. The Hall conductivity breaking TRI is therefore allowed, since $\sigma_{xy}$ is also a property of the fermionic subsystem only. 

One question left unanswered is how the lack of TRI of the fermionic subsystem is balanced by a corresponding TRI breaking in the bosonic degrees of freedom, so that the system as a whole remains TR invariant. For real scalar bosons, the only possible observables which are related to breaking of TRI are odd correlation functions in $\phi_q$. Therefore, one could in principle see the compensating TR breaking in the boson sector by computing such observables. More seriously, since the bosons are chargeless, the TR-breaking Hall current has to be compensated for by a current carried by the fermions in the reservoir. A complete explanation therefore needs to take the fermions exiting the system via \eqref{jump} into account, together with their interactions with the EM field and the bosonic sector. A detailed description of this is beyond the scope of this work. Another interesting direction for future research is to connect our results more systematically to the general understanding of quantum field theory anomalies.

For a possible experimental realization, we note that our system consists of fermions confined to a flat material sample and exposed to a bosonic bath. This is generic, and the bosons could be e.g. phonons in the sample. Since $m=0$ in the boson dispersion \eqref{dispbos}, and $c_s$ is taken very small, the bosons would then be acoustic phonons in a material with very low speed of sound. The most challenging part to implement is the jump operators \eqref{jump}, which allow fermions to enter or leave the system with a specific modulation by the bosons in a way that introduces chirality to the system. A possible mechanism is that fermions collide with bosons and are pushed into or out of the system as a result, but the specific details are nontrivial. Therefore, the jump operators, used to motivate the dissipative action \eqref{SD} which is our starting point, are the main limitation to the universality of our results.

\section{Acknowledgments}
The authors wish to thank Eddy Ardonne, Emil J. Bergholtz, and T. Hans Hansson for illuminating discussions and for their support during this project. RA is also grateful for the reception and hospitality in the KOMKO corridor in Fysikum during the early stages of this project. CE acknowledges financial support by the G\"oran Gustafsson Foundation for Research in Natural Sciences and Medicine. RA acknowledges the support of the INCT project
Advanced Quantum Materials, involving the Brazilian agencies CNPq (Proc.
408766/2024-7), FAPESP (Proc. 2025/27091-3), and CAPES.
\bibliography{refs}

\begin{thebibliography}{79}%
\makeatletter
\providecommand \@ifxundefined [1]{%
 \@ifx{#1\undefined}
}%
\providecommand \@ifnum [1]{%
 \ifnum #1\expandafter \@firstoftwo
 \else \expandafter \@secondoftwo
 \fi
}%
\providecommand \@ifx [1]{%
 \ifx #1\expandafter \@firstoftwo
 \else \expandafter \@secondoftwo
 \fi
}%
\providecommand \natexlab [1]{#1}%
\providecommand \enquote  [1]{``#1''}%
\providecommand \bibnamefont  [1]{#1}%
\providecommand \bibfnamefont [1]{#1}%
\providecommand \citenamefont [1]{#1}%
\providecommand \href@noop [0]{\@secondoftwo}%
\providecommand \href [0]{\begingroup \@sanitize@url \@href}%
\providecommand \@href[1]{\@@startlink{#1}\@@href}%
\providecommand \@@href[1]{\endgroup#1\@@endlink}%
\providecommand \@sanitize@url [0]{\catcode `\\12\catcode `\$12\catcode `\&12\catcode `\#12\catcode `\^12\catcode `\_12\catcode `\%12\relax}%
\providecommand \@@startlink[1]{}%
\providecommand \@@endlink[0]{}%
\providecommand \url  [0]{\begingroup\@sanitize@url \@url }%
\providecommand \@url [1]{\endgroup\@href {#1}{\urlprefix }}%
\providecommand \urlprefix  [0]{URL }%
\providecommand \Eprint [0]{\href }%
\providecommand \doibase [0]{https://doi.org/}%
\providecommand \selectlanguage [0]{\@gobble}%
\providecommand \bibinfo  [0]{\@secondoftwo}%
\providecommand \bibfield  [0]{\@secondoftwo}%
\providecommand \translation [1]{[#1]}%
\providecommand \BibitemOpen [0]{}%
\providecommand \bibitemStop [0]{}%
\providecommand \bibitemNoStop [0]{.\EOS\space}%
\providecommand \EOS [0]{\spacefactor3000\relax}%
\providecommand \BibitemShut  [1]{\csname bibitem#1\endcsname}%
\let\auto@bib@innerbib\@empty
\bibitem [{\citenamefont {Klitzing}\ \emph {et~al.}(1980)\citenamefont {Klitzing}, \citenamefont {Dorda},\ and\ \citenamefont {Pepper}}]{klitzing1980qhe}%
  \BibitemOpen
  \bibfield  {author} {\bibinfo {author} {\bibfnamefont {K.~v.}\ \bibnamefont {Klitzing}}, \bibinfo {author} {\bibfnamefont {G.}~\bibnamefont {Dorda}},\ and\ \bibinfo {author} {\bibfnamefont {M.}~\bibnamefont {Pepper}},\ }\bibfield  {title} {\bibinfo {title} {New method for high-accuracy determination of the fine-structure constant based on quantized hall resistance},\ }\href {https://doi.org/10.1103/PhysRevLett.45.494} {\bibfield  {journal} {\bibinfo  {journal} {Phys. Rev. Lett.}\ }\textbf {\bibinfo {volume} {45}},\ \bibinfo {pages} {494} (\bibinfo {year} {1980})}\BibitemShut {NoStop}%
\bibitem [{\citenamefont {Tsui}\ \emph {et~al.}(1982)\citenamefont {Tsui}, \citenamefont {Stormer},\ and\ \citenamefont {Gossard}}]{tsui1982fqhe}%
  \BibitemOpen
  \bibfield  {author} {\bibinfo {author} {\bibfnamefont {D.~C.}\ \bibnamefont {Tsui}}, \bibinfo {author} {\bibfnamefont {H.~L.}\ \bibnamefont {Stormer}},\ and\ \bibinfo {author} {\bibfnamefont {A.~C.}\ \bibnamefont {Gossard}},\ }\bibfield  {title} {\bibinfo {title} {Two-dimensional magnetotransport in the extreme quantum limit},\ }\href {https://doi.org/10.1103/PhysRevLett.48.1559} {\bibfield  {journal} {\bibinfo  {journal} {Phys. Rev. Lett.}\ }\textbf {\bibinfo {volume} {48}},\ \bibinfo {pages} {1559} (\bibinfo {year} {1982})}\BibitemShut {NoStop}%
\bibitem [{\citenamefont {Wilczek}(1982)}]{wilczek1982fractional}%
  \BibitemOpen
  \bibfield  {author} {\bibinfo {author} {\bibfnamefont {F.}~\bibnamefont {Wilczek}},\ }\bibfield  {title} {\bibinfo {title} {Quantum mechanics of fractional-spin particles},\ }\href {https://doi.org/10.1103/PhysRevLett.49.957} {\bibfield  {journal} {\bibinfo  {journal} {Phys. Rev. Lett.}\ }\textbf {\bibinfo {volume} {49}},\ \bibinfo {pages} {957} (\bibinfo {year} {1982})}\BibitemShut {NoStop}%
\bibitem [{\citenamefont {Thouless}\ \emph {et~al.}(1982)\citenamefont {Thouless}, \citenamefont {Kohmoto}, \citenamefont {Nightingale},\ and\ \citenamefont {den Nijs}}]{thouless1982quantized}%
  \BibitemOpen
  \bibfield  {author} {\bibinfo {author} {\bibfnamefont {D.~J.}\ \bibnamefont {Thouless}}, \bibinfo {author} {\bibfnamefont {M.}~\bibnamefont {Kohmoto}}, \bibinfo {author} {\bibfnamefont {M.~P.}\ \bibnamefont {Nightingale}},\ and\ \bibinfo {author} {\bibfnamefont {M.}~\bibnamefont {den Nijs}},\ }\bibfield  {title} {\bibinfo {title} {Quantized hall conductance in a two-dimensional periodic potential},\ }\href {https://doi.org/10.1103/PhysRevLett.49.405} {\bibfield  {journal} {\bibinfo  {journal} {Phys. Rev. Lett.}\ }\textbf {\bibinfo {volume} {49}},\ \bibinfo {pages} {405} (\bibinfo {year} {1982})}\BibitemShut {NoStop}%
\bibitem [{\citenamefont {Halperin}(1984)}]{halperin1984statistics}%
  \BibitemOpen
  \bibfield  {author} {\bibinfo {author} {\bibfnamefont {B.~I.}\ \bibnamefont {Halperin}},\ }\bibfield  {title} {\bibinfo {title} {Statistics of quasiparticles and the hierarchy of fractional quantized hall states},\ }\href {https://doi.org/10.1103/PhysRevLett.52.1583} {\bibfield  {journal} {\bibinfo  {journal} {Phys. Rev. Lett.}\ }\textbf {\bibinfo {volume} {52}},\ \bibinfo {pages} {1583} (\bibinfo {year} {1984})}\BibitemShut {NoStop}%
\bibitem [{\citenamefont {Arovas}\ \emph {et~al.}(1984)\citenamefont {Arovas}, \citenamefont {Schrieffer},\ and\ \citenamefont {Wilczek}}]{arovas1984statistics}%
  \BibitemOpen
  \bibfield  {author} {\bibinfo {author} {\bibfnamefont {D.}~\bibnamefont {Arovas}}, \bibinfo {author} {\bibfnamefont {J.~R.}\ \bibnamefont {Schrieffer}},\ and\ \bibinfo {author} {\bibfnamefont {F.}~\bibnamefont {Wilczek}},\ }\bibfield  {title} {\bibinfo {title} {Fractional statistics and the quantum hall effect},\ }\href {https://doi.org/10.1103/PhysRevLett.53.722} {\bibfield  {journal} {\bibinfo  {journal} {Phys. Rev. Lett.}\ }\textbf {\bibinfo {volume} {53}},\ \bibinfo {pages} {722} (\bibinfo {year} {1984})}\BibitemShut {NoStop}%
\bibitem [{\citenamefont {Girvin}\ and\ \citenamefont {MacDonald}(1987)}]{girvin1987order}%
  \BibitemOpen
  \bibfield  {author} {\bibinfo {author} {\bibfnamefont {S.~M.}\ \bibnamefont {Girvin}}\ and\ \bibinfo {author} {\bibfnamefont {A.~H.}\ \bibnamefont {MacDonald}},\ }\bibfield  {title} {\bibinfo {title} {Off-diagonal long-range order, oblique confinement, and the fractional quantum hall effect},\ }\href {https://doi.org/10.1103/PhysRevLett.58.1252} {\bibfield  {journal} {\bibinfo  {journal} {Phys. Rev. Lett.}\ }\textbf {\bibinfo {volume} {58}},\ \bibinfo {pages} {1252} (\bibinfo {year} {1987})}\BibitemShut {NoStop}%
\bibitem [{\citenamefont {Zhang}\ \emph {et~al.}(1989)\citenamefont {Zhang}, \citenamefont {Hansson},\ and\ \citenamefont {Kivelson}}]{zhang1989eft}%
  \BibitemOpen
  \bibfield  {author} {\bibinfo {author} {\bibfnamefont {S.~C.}\ \bibnamefont {Zhang}}, \bibinfo {author} {\bibfnamefont {T.~H.}\ \bibnamefont {Hansson}},\ and\ \bibinfo {author} {\bibfnamefont {S.}~\bibnamefont {Kivelson}},\ }\bibfield  {title} {\bibinfo {title} {Effective-field-theory model for the fractional quantum hall effect},\ }\href {https://doi.org/10.1103/PhysRevLett.62.82} {\bibfield  {journal} {\bibinfo  {journal} {Phys. Rev. Lett.}\ }\textbf {\bibinfo {volume} {62}},\ \bibinfo {pages} {82} (\bibinfo {year} {1989})}\BibitemShut {NoStop}%
\bibitem [{\citenamefont {Goldman}\ and\ \citenamefont {Su}(1995)}]{goldman1995tunneling}%
  \BibitemOpen
  \bibfield  {author} {\bibinfo {author} {\bibfnamefont {V.~J.}\ \bibnamefont {Goldman}}\ and\ \bibinfo {author} {\bibfnamefont {B.}~\bibnamefont {Su}},\ }\bibfield  {title} {\bibinfo {title} {Resonant tunneling in the quantum hall regime: Measurement of fractional charge},\ }\href {https://doi.org/10.1126/science.267.5200.1010} {\bibfield  {journal} {\bibinfo  {journal} {Science}\ }\textbf {\bibinfo {volume} {267}},\ \bibinfo {pages} {1010} (\bibinfo {year} {1995})},\ \Eprint {https://arxiv.org/abs/https://www.science.org/doi/pdf/10.1126/science.267.5200.1010} {https://www.science.org/doi/pdf/10.1126/science.267.5200.1010} \BibitemShut {NoStop}%
\bibitem [{\citenamefont {Saminadayar}\ \emph {et~al.}(1997)\citenamefont {Saminadayar}, \citenamefont {Glattli}, \citenamefont {Jin},\ and\ \citenamefont {Etienne}}]{saminadayar1997charge}%
  \BibitemOpen
  \bibfield  {author} {\bibinfo {author} {\bibfnamefont {L.}~\bibnamefont {Saminadayar}}, \bibinfo {author} {\bibfnamefont {D.~C.}\ \bibnamefont {Glattli}}, \bibinfo {author} {\bibfnamefont {Y.}~\bibnamefont {Jin}},\ and\ \bibinfo {author} {\bibfnamefont {B.}~\bibnamefont {Etienne}},\ }\bibfield  {title} {\bibinfo {title} {Observation of the $\mathit{e}\mathit{/}3$ fractionally charged laughlin quasiparticle},\ }\href {https://doi.org/10.1103/PhysRevLett.79.2526} {\bibfield  {journal} {\bibinfo  {journal} {Phys. Rev. Lett.}\ }\textbf {\bibinfo {volume} {79}},\ \bibinfo {pages} {2526} (\bibinfo {year} {1997})}\BibitemShut {NoStop}%
\bibitem [{\citenamefont {Nakamura}\ \emph {et~al.}(2020)\citenamefont {Nakamura}, \citenamefont {Liang}, \citenamefont {Gardner},\ and\ \citenamefont {Manfra}}]{nakamura2020observation}%
  \BibitemOpen
  \bibfield  {author} {\bibinfo {author} {\bibfnamefont {J.}~\bibnamefont {Nakamura}}, \bibinfo {author} {\bibfnamefont {S.}~\bibnamefont {Liang}}, \bibinfo {author} {\bibfnamefont {G.~C.}\ \bibnamefont {Gardner}},\ and\ \bibinfo {author} {\bibfnamefont {M.~J.}\ \bibnamefont {Manfra}},\ }\bibfield  {title} {\bibinfo {title} {Direct observation of anyonic braiding statistics},\ }\bibfield  {journal} {\bibinfo  {journal} {Nature Physics}\ }\textbf {\bibinfo {volume} {16}},\ \href {https://doi.org/10.1038/s41567-020-1019-1} {10.1038/s41567-020-1019-1} (\bibinfo {year} {2020})\BibitemShut {NoStop}%
\bibitem [{\citenamefont {Comparin}\ \emph {et~al.}(2022)\citenamefont {Comparin}, \citenamefont {Opler}, \citenamefont {Macaluso}, \citenamefont {Biella}, \citenamefont {Polychronakos},\ and\ \citenamefont {Mazza}}]{comparin2022spin}%
  \BibitemOpen
  \bibfield  {author} {\bibinfo {author} {\bibfnamefont {T.}~\bibnamefont {Comparin}}, \bibinfo {author} {\bibfnamefont {A.}~\bibnamefont {Opler}}, \bibinfo {author} {\bibfnamefont {E.}~\bibnamefont {Macaluso}}, \bibinfo {author} {\bibfnamefont {A.}~\bibnamefont {Biella}}, \bibinfo {author} {\bibfnamefont {A.~P.}\ \bibnamefont {Polychronakos}},\ and\ \bibinfo {author} {\bibfnamefont {L.}~\bibnamefont {Mazza}},\ }\bibfield  {title} {\bibinfo {title} {Measurable fractional spin for quantum hall quasiparticles on the disk},\ }\href {https://doi.org/10.1103/PhysRevB.105.085125} {\bibfield  {journal} {\bibinfo  {journal} {Phys. Rev. B}\ }\textbf {\bibinfo {volume} {105}},\ \bibinfo {pages} {085125} (\bibinfo {year} {2022})}\BibitemShut {NoStop}%
\bibitem [{\citenamefont {Nardin}\ \emph {et~al.}(2023)\citenamefont {Nardin}, \citenamefont {Ardonne},\ and\ \citenamefont {Mazza}}]{nardin2023ssr}%
  \BibitemOpen
  \bibfield  {author} {\bibinfo {author} {\bibfnamefont {A.}~\bibnamefont {Nardin}}, \bibinfo {author} {\bibfnamefont {E.}~\bibnamefont {Ardonne}},\ and\ \bibinfo {author} {\bibfnamefont {L.}~\bibnamefont {Mazza}},\ }\bibfield  {title} {\bibinfo {title} {Spin-statistics relation for quantum hall states},\ }\href {https://doi.org/10.1103/PhysRevB.108.L041105} {\bibfield  {journal} {\bibinfo  {journal} {Phys. Rev. B}\ }\textbf {\bibinfo {volume} {108}},\ \bibinfo {pages} {L041105} (\bibinfo {year} {2023})}\BibitemShut {NoStop}%
\bibitem [{\citenamefont {Fagerlund}\ \emph {et~al.}(2025)\citenamefont {Fagerlund}, \citenamefont {Nardin}, \citenamefont {Mazza},\ and\ \citenamefont {Ardonne}}]{fagerlund2025spin}%
  \BibitemOpen
  \bibfield  {author} {\bibinfo {author} {\bibfnamefont {A.}~\bibnamefont {Fagerlund}}, \bibinfo {author} {\bibfnamefont {A.}~\bibnamefont {Nardin}}, \bibinfo {author} {\bibfnamefont {L.}~\bibnamefont {Mazza}},\ and\ \bibinfo {author} {\bibfnamefont {E.}~\bibnamefont {Ardonne}},\ }\bibfield  {title} {\bibinfo {title} {Spin fractionalization at the edge of quantum hall fluids induced by bulk quasiparticles},\ }\href {https://doi.org/10.1103/t5ww-65lz} {\bibfield  {journal} {\bibinfo  {journal} {Phys. Rev. B}\ }\textbf {\bibinfo {volume} {112}},\ \bibinfo {pages} {075148} (\bibinfo {year} {2025})}\BibitemShut {NoStop}%
\bibitem [{\citenamefont {Fagerlund}\ and\ \citenamefont {Ardonne}(2025)}]{fagerlund2025bosonic}%
  \BibitemOpen
  \bibfield  {author} {\bibinfo {author} {\bibfnamefont {A.}~\bibnamefont {Fagerlund}}\ and\ \bibinfo {author} {\bibfnamefont {E.}~\bibnamefont {Ardonne}},\ }\bibfield  {title} {\bibinfo {title} {Bosonic matrix product state description of read-rezayi states and its application to quasihole spins},\ }\href {https://doi.org/10.1103/pgsb-wjts} {\bibfield  {journal} {\bibinfo  {journal} {Phys. Rev. B}\ }\textbf {\bibinfo {volume} {112}},\ \bibinfo {pages} {125126} (\bibinfo {year} {2025})}\BibitemShut {NoStop}%
\bibitem [{\citenamefont {Moore}\ and\ \citenamefont {Read}(1991)}]{moore1991nonabelions}%
  \BibitemOpen
  \bibfield  {author} {\bibinfo {author} {\bibfnamefont {G.}~\bibnamefont {Moore}}\ and\ \bibinfo {author} {\bibfnamefont {N.}~\bibnamefont {Read}},\ }\bibfield  {title} {\bibinfo {title} {Nonabelions in the fractional quantum hall effect},\ }\href {https://doi.org/https://doi.org/10.1016/0550-3213(91)90407-O} {\bibfield  {journal} {\bibinfo  {journal} {Nuclear Physics B}\ }\textbf {\bibinfo {volume} {360}},\ \bibinfo {pages} {362} (\bibinfo {year} {1991})}\BibitemShut {NoStop}%
\bibitem [{\citenamefont {Kitaev}(2003)}]{kitaev2003computation}%
  \BibitemOpen
  \bibfield  {author} {\bibinfo {author} {\bibfnamefont {A.}~\bibnamefont {Kitaev}},\ }\bibfield  {title} {\bibinfo {title} {Fault-tolerant quantum computation by anyons},\ }\href {https://doi.org/https://doi.org/10.1016/S0003-4916(02)00018-0} {\bibfield  {journal} {\bibinfo  {journal} {Annals of Physics}\ }\textbf {\bibinfo {volume} {303}},\ \bibinfo {pages} {2} (\bibinfo {year} {2003})}\BibitemShut {NoStop}%
\bibitem [{\citenamefont {Nayak}\ \emph {et~al.}(2008)\citenamefont {Nayak}, \citenamefont {Simon}, \citenamefont {Stern}, \citenamefont {Freedman},\ and\ \citenamefont {Das~Sarma}}]{nayak2008nonabelian}%
  \BibitemOpen
  \bibfield  {author} {\bibinfo {author} {\bibfnamefont {C.}~\bibnamefont {Nayak}}, \bibinfo {author} {\bibfnamefont {S.~H.}\ \bibnamefont {Simon}}, \bibinfo {author} {\bibfnamefont {A.}~\bibnamefont {Stern}}, \bibinfo {author} {\bibfnamefont {M.}~\bibnamefont {Freedman}},\ and\ \bibinfo {author} {\bibfnamefont {S.}~\bibnamefont {Das~Sarma}},\ }\bibfield  {title} {\bibinfo {title} {Non-abelian anyons and topological quantum computation},\ }\href@noop {} {\bibfield  {journal} {\bibinfo  {journal} {Reviews of Modern Physics}\ }\textbf {\bibinfo {volume} {80}},\ \bibinfo {pages} {1083} (\bibinfo {year} {2008})}\BibitemShut {NoStop}%
\bibitem [{\citenamefont {Altland}\ and\ \citenamefont {Zirnbauer}(1997)}]{altland1997nonstandard}%
  \BibitemOpen
  \bibfield  {author} {\bibinfo {author} {\bibfnamefont {A.}~\bibnamefont {Altland}}\ and\ \bibinfo {author} {\bibfnamefont {M.~R.}\ \bibnamefont {Zirnbauer}},\ }\bibfield  {title} {\bibinfo {title} {Nonstandard symmetry classes in mesoscopic normal-superconducting hybrid structures},\ }\href {https://doi.org/10.1103/PhysRevB.55.1142} {\bibfield  {journal} {\bibinfo  {journal} {Phys. Rev. B}\ }\textbf {\bibinfo {volume} {55}},\ \bibinfo {pages} {1142} (\bibinfo {year} {1997})}\BibitemShut {NoStop}%
\bibitem [{\citenamefont {Schnyder}\ \emph {et~al.}(2008)\citenamefont {Schnyder}, \citenamefont {Ryu}, \citenamefont {Furusaki},\ and\ \citenamefont {Ludwig}}]{schnyder2008classification}%
  \BibitemOpen
  \bibfield  {author} {\bibinfo {author} {\bibfnamefont {A.~P.}\ \bibnamefont {Schnyder}}, \bibinfo {author} {\bibfnamefont {S.}~\bibnamefont {Ryu}}, \bibinfo {author} {\bibfnamefont {A.}~\bibnamefont {Furusaki}},\ and\ \bibinfo {author} {\bibfnamefont {A.~W.~W.}\ \bibnamefont {Ludwig}},\ }\bibfield  {title} {\bibinfo {title} {Classification of topological insulators and superconductors in three spatial dimensions},\ }\href {https://doi.org/10.1103/PhysRevB.78.195125} {\bibfield  {journal} {\bibinfo  {journal} {Phys. Rev. B}\ }\textbf {\bibinfo {volume} {78}},\ \bibinfo {pages} {195125} (\bibinfo {year} {2008})}\BibitemShut {NoStop}%
\bibitem [{\citenamefont {Kitaev}(2009)}]{kitaev2009table}%
  \BibitemOpen
  \bibfield  {author} {\bibinfo {author} {\bibfnamefont {A.}~\bibnamefont {Kitaev}},\ }\bibfield  {title} {\bibinfo {title} {Periodic table for topological insulators and superconductors},\ }\href {https://doi.org/10.1063/1.3149495} {\bibfield  {journal} {\bibinfo  {journal} {AIP Conference Proceedings}\ }\textbf {\bibinfo {volume} {1134}},\ \bibinfo {pages} {22} (\bibinfo {year} {2009})}\BibitemShut {NoStop}%
\bibitem [{\citenamefont {Ryu}\ \emph {et~al.}(2010)\citenamefont {Ryu}, \citenamefont {Schnyder}, \citenamefont {Furusaki},\ and\ \citenamefont {Ludwig}}]{ryu2010topological}%
  \BibitemOpen
  \bibfield  {author} {\bibinfo {author} {\bibfnamefont {S.}~\bibnamefont {Ryu}}, \bibinfo {author} {\bibfnamefont {A.~P.}\ \bibnamefont {Schnyder}}, \bibinfo {author} {\bibfnamefont {A.}~\bibnamefont {Furusaki}},\ and\ \bibinfo {author} {\bibfnamefont {A.~W.}\ \bibnamefont {Ludwig}},\ }\bibfield  {title} {\bibinfo {title} {Topological insulators and superconductors: tenfold way and dimensional hierarchy},\ }\href@noop {} {\bibfield  {journal} {\bibinfo  {journal} {New Journal of Physics}\ }\textbf {\bibinfo {volume} {12}},\ \bibinfo {pages} {065010} (\bibinfo {year} {2010})}\BibitemShut {NoStop}%
\bibitem [{\citenamefont {Wen}(2017)}]{Wen2017RMP}%
  \BibitemOpen
  \bibfield  {author} {\bibinfo {author} {\bibfnamefont {X.-G.}\ \bibnamefont {Wen}},\ }\bibfield  {title} {\bibinfo {title} {Colloquium: Zoo of quantum-topological phases of matter},\ }\href {https://doi.org/10.1103/RevModPhys.89.041004} {\bibfield  {journal} {\bibinfo  {journal} {Rev. Mod. Phys.}\ }\textbf {\bibinfo {volume} {89}},\ \bibinfo {pages} {041004} (\bibinfo {year} {2017})}\BibitemShut {NoStop}%
\bibitem [{\citenamefont {Fu}(2011)}]{fu2011topological}%
  \BibitemOpen
  \bibfield  {author} {\bibinfo {author} {\bibfnamefont {L.}~\bibnamefont {Fu}},\ }\bibfield  {title} {\bibinfo {title} {Topological crystalline insulators},\ }\href {https://doi.org/10.1103/PhysRevLett.106.106802} {\bibfield  {journal} {\bibinfo  {journal} {Phys. Rev. Lett.}\ }\textbf {\bibinfo {volume} {106}},\ \bibinfo {pages} {106802} (\bibinfo {year} {2011})}\BibitemShut {NoStop}%
\bibitem [{\citenamefont {Slager}\ \emph {et~al.}(2013)\citenamefont {Slager}, \citenamefont {Mesaros}, \citenamefont {Juri{\v{c}}i{\'{c}}},\ and\ \citenamefont {Zaanen}}]{slager2013space}%
  \BibitemOpen
  \bibfield  {author} {\bibinfo {author} {\bibfnamefont {R.-J.}\ \bibnamefont {Slager}}, \bibinfo {author} {\bibfnamefont {A.}~\bibnamefont {Mesaros}}, \bibinfo {author} {\bibfnamefont {V.}~\bibnamefont {Juri{\v{c}}i{\'{c}}}},\ and\ \bibinfo {author} {\bibfnamefont {J.}~\bibnamefont {Zaanen}},\ }\bibfield  {title} {\bibinfo {title} {The space group classification of topological band-insulators},\ }\href {https://doi.org/10.1038/nphys2513} {\bibfield  {journal} {\bibinfo  {journal} {Nature Physics}\ }\textbf {\bibinfo {volume} {9}},\ \bibinfo {pages} {98} (\bibinfo {year} {2013})}\BibitemShut {NoStop}%
\bibitem [{\citenamefont {Kitagawa}\ \emph {et~al.}(2010)\citenamefont {Kitagawa}, \citenamefont {Berg}, \citenamefont {Rudner},\ and\ \citenamefont {Demler}}]{Kitagawa2010PRB}%
  \BibitemOpen
  \bibfield  {author} {\bibinfo {author} {\bibfnamefont {T.}~\bibnamefont {Kitagawa}}, \bibinfo {author} {\bibfnamefont {E.}~\bibnamefont {Berg}}, \bibinfo {author} {\bibfnamefont {M.~S.}\ \bibnamefont {Rudner}},\ and\ \bibinfo {author} {\bibfnamefont {E.}~\bibnamefont {Demler}},\ }\bibfield  {title} {\bibinfo {title} {Topological characterization of periodically driven quantum systems},\ }\href {https://doi.org/10.1103/PhysRevB.82.235114} {\bibfield  {journal} {\bibinfo  {journal} {Physical Review B}\ }\textbf {\bibinfo {volume} {82}},\ \bibinfo {pages} {235114} (\bibinfo {year} {2010})}\BibitemShut {NoStop}%
\bibitem [{\citenamefont {Lindner}\ \emph {et~al.}(2011)\citenamefont {Lindner}, \citenamefont {Refael},\ and\ \citenamefont {Galitski}}]{Lindner2011NatPhys}%
  \BibitemOpen
  \bibfield  {author} {\bibinfo {author} {\bibfnamefont {N.~H.}\ \bibnamefont {Lindner}}, \bibinfo {author} {\bibfnamefont {G.}~\bibnamefont {Refael}},\ and\ \bibinfo {author} {\bibfnamefont {V.}~\bibnamefont {Galitski}},\ }\bibfield  {title} {\bibinfo {title} {Floquet topological insulator in semiconductor quantum wells},\ }\href {https://doi.org/10.1038/nphys1926} {\bibfield  {journal} {\bibinfo  {journal} {Nature Physics}\ }\textbf {\bibinfo {volume} {7}},\ \bibinfo {pages} {490} (\bibinfo {year} {2011})}\BibitemShut {NoStop}%
\bibitem [{\citenamefont {Ashida}\ \emph {et~al.}(2020)\citenamefont {Ashida}, \citenamefont {Gong},\ and\ \citenamefont {Ueda}}]{Ashida2020AP}%
  \BibitemOpen
  \bibfield  {author} {\bibinfo {author} {\bibfnamefont {Y.}~\bibnamefont {Ashida}}, \bibinfo {author} {\bibfnamefont {Z.}~\bibnamefont {Gong}},\ and\ \bibinfo {author} {\bibfnamefont {M.}~\bibnamefont {Ueda}},\ }\bibfield  {title} {\bibinfo {title} {Non-hermitian physics},\ }\href {https://doi.org/10.1080/00018732.2021.1876991} {\bibfield  {journal} {\bibinfo  {journal} {Adv. Phys.}\ }\textbf {\bibinfo {volume} {69}},\ \bibinfo {pages} {249} (\bibinfo {year} {2020})}\BibitemShut {NoStop}%
\bibitem [{\citenamefont {Bergholtz}\ \emph {et~al.}(2021)\citenamefont {Bergholtz}, \citenamefont {Budich},\ and\ \citenamefont {Kunst}}]{Bergholtz2021RMP}%
  \BibitemOpen
  \bibfield  {author} {\bibinfo {author} {\bibfnamefont {E.~J.}\ \bibnamefont {Bergholtz}}, \bibinfo {author} {\bibfnamefont {J.~C.}\ \bibnamefont {Budich}},\ and\ \bibinfo {author} {\bibfnamefont {F.~K.}\ \bibnamefont {Kunst}},\ }\bibfield  {title} {\bibinfo {title} {Exceptional topology of non-hermitian systems},\ }\href {https://doi.org/10.1103/RevModPhys.93.015005} {\bibfield  {journal} {\bibinfo  {journal} {Rev. Mod. Phys.}\ }\textbf {\bibinfo {volume} {93}},\ \bibinfo {pages} {015005} (\bibinfo {year} {2021})}\BibitemShut {NoStop}%
\bibitem [{\citenamefont {Okuma}\ and\ \citenamefont {Sato}(2023)}]{Okuma2023ARCMP}%
  \BibitemOpen
  \bibfield  {author} {\bibinfo {author} {\bibfnamefont {N.}~\bibnamefont {Okuma}}\ and\ \bibinfo {author} {\bibfnamefont {M.}~\bibnamefont {Sato}},\ }\bibfield  {title} {\bibinfo {title} {Non-hermitian topological phenomena: A review},\ }\href {https://doi.org/10.1146/annurev-conmatphys-040521-033133} {\bibfield  {journal} {\bibinfo  {journal} {Annu. Rev. Condens. Matter Phys}\ }\textbf {\bibinfo {volume} {14}},\ \bibinfo {pages} {83} (\bibinfo {year} {2023})}\BibitemShut {NoStop}%
\bibitem [{\citenamefont {Feshbach}(1962)}]{Feshbach1962AnnPhys}%
  \BibitemOpen
  \bibfield  {author} {\bibinfo {author} {\bibfnamefont {H.}~\bibnamefont {Feshbach}},\ }\bibfield  {title} {\bibinfo {title} {A unified theory of nuclear reactions. ii},\ }\href {https://doi.org/10.1016/0003-4916(62)90221-X} {\bibfield  {journal} {\bibinfo  {journal} {Annals of Physics}\ }\textbf {\bibinfo {volume} {19}},\ \bibinfo {pages} {287} (\bibinfo {year} {1962})}\BibitemShut {NoStop}%
\bibitem [{\citenamefont {Dalibard}\ \emph {et~al.}(1992)\citenamefont {Dalibard}, \citenamefont {Castin},\ and\ \citenamefont {M{\o}lmer}}]{Dalibard1992PRL}%
  \BibitemOpen
  \bibfield  {author} {\bibinfo {author} {\bibfnamefont {J.}~\bibnamefont {Dalibard}}, \bibinfo {author} {\bibfnamefont {Y.}~\bibnamefont {Castin}},\ and\ \bibinfo {author} {\bibfnamefont {K.}~\bibnamefont {M{\o}lmer}},\ }\bibfield  {title} {\bibinfo {title} {Wave-function approach to dissipative processes in quantum optics},\ }\href {https://doi.org/10.1103/PhysRevLett.68.580} {\bibfield  {journal} {\bibinfo  {journal} {Physical Review Letters}\ }\textbf {\bibinfo {volume} {68}},\ \bibinfo {pages} {580} (\bibinfo {year} {1992})}\BibitemShut {NoStop}%
\bibitem [{\citenamefont {Dum}\ \emph {et~al.}(1992)\citenamefont {Dum}, \citenamefont {Zoller},\ and\ \citenamefont {Ritsch}}]{Dum1992PRA}%
  \BibitemOpen
  \bibfield  {author} {\bibinfo {author} {\bibfnamefont {R.}~\bibnamefont {Dum}}, \bibinfo {author} {\bibfnamefont {P.}~\bibnamefont {Zoller}},\ and\ \bibinfo {author} {\bibfnamefont {H.}~\bibnamefont {Ritsch}},\ }\bibfield  {title} {\bibinfo {title} {Monte carlo simulation of the atomic master equation for spontaneous emission},\ }\href {https://doi.org/10.1103/PhysRevA.45.4879} {\bibfield  {journal} {\bibinfo  {journal} {Physical Review A}\ }\textbf {\bibinfo {volume} {45}},\ \bibinfo {pages} {4879} (\bibinfo {year} {1992})}\BibitemShut {NoStop}%
\bibitem [{\citenamefont {Breuer}\ and\ \citenamefont {Petruccione}(2002)}]{BreuerPetruccione2002}%
  \BibitemOpen
  \bibfield  {author} {\bibinfo {author} {\bibfnamefont {H.-P.}\ \bibnamefont {Breuer}}\ and\ \bibinfo {author} {\bibfnamefont {F.}~\bibnamefont {Petruccione}},\ }\href@noop {} {\emph {\bibinfo {title} {The Theory of Open Quantum Systems}}}\ (\bibinfo  {publisher} {Oxford University Press},\ \bibinfo {year} {2002})\BibitemShut {NoStop}%
\bibitem [{\citenamefont {Prosen}(2008)}]{Prosen2008NJP}%
  \BibitemOpen
  \bibfield  {author} {\bibinfo {author} {\bibfnamefont {T.}~\bibnamefont {Prosen}},\ }\bibfield  {title} {\bibinfo {title} {Third quantization: a general method to solve master equations for quadratic open fermi systems},\ }\href {https://doi.org/10.1088/1367-2630/10/4/043026} {\bibfield  {journal} {\bibinfo  {journal} {New Journal of Physics}\ }\textbf {\bibinfo {volume} {10}},\ \bibinfo {pages} {043026} (\bibinfo {year} {2008})}\BibitemShut {NoStop}%
\bibitem [{\citenamefont {Rotter}(2009)}]{Rotter2009JPA}%
  \BibitemOpen
  \bibfield  {author} {\bibinfo {author} {\bibfnamefont {I.}~\bibnamefont {Rotter}},\ }\bibfield  {title} {\bibinfo {title} {A non-hermitian hamilton operator and the physics of open quantum systems},\ }\href {https://doi.org/10.1088/1751-8113/42/15/153001} {\bibfield  {journal} {\bibinfo  {journal} {Journal of Physics A: Mathematical and Theoretical}\ }\textbf {\bibinfo {volume} {42}},\ \bibinfo {pages} {153001} (\bibinfo {year} {2009})}\BibitemShut {NoStop}%
\bibitem [{\citenamefont {Ekman}\ and\ \citenamefont {Bergholtz}(2024)}]{Ekman2024PRR}%
  \BibitemOpen
  \bibfield  {author} {\bibinfo {author} {\bibfnamefont {C.}~\bibnamefont {Ekman}}\ and\ \bibinfo {author} {\bibfnamefont {E.~J.}\ \bibnamefont {Bergholtz}},\ }\bibfield  {title} {\bibinfo {title} {Liouvillian skin effects and fragmented condensates in an integrable dissipative bose-hubbard model},\ }\href {https://doi.org/10.1103/PhysRevResearch.6.L032067} {\bibfield  {journal} {\bibinfo  {journal} {Phys. Rev. Res.}\ }\textbf {\bibinfo {volume} {6}},\ \bibinfo {pages} {L032067} (\bibinfo {year} {2024})}\BibitemShut {NoStop}%
\bibitem [{\citenamefont {Kozii}\ and\ \citenamefont {Fu}(2017)}]{Kozii2017PRB}%
  \BibitemOpen
  \bibfield  {author} {\bibinfo {author} {\bibfnamefont {V.}~\bibnamefont {Kozii}}\ and\ \bibinfo {author} {\bibfnamefont {L.}~\bibnamefont {Fu}},\ }\bibfield  {title} {\bibinfo {title} {Non-hermitian topological theory of finite-lifetime quasiparticles: Prediction of bulk fermi arc due to exceptional point},\ }\href {https://doi.org/10.1103/PhysRevB.96.155138} {\bibfield  {journal} {\bibinfo  {journal} {Physical Review B}\ }\textbf {\bibinfo {volume} {96}},\ \bibinfo {pages} {155138} (\bibinfo {year} {2017})}\BibitemShut {NoStop}%
\bibitem [{\citenamefont {Zyuzin}\ \emph {et~al.}(2018)\citenamefont {Zyuzin}, \citenamefont {Wu},\ and\ \citenamefont {Burkov}}]{Zyuzin2018PRB}%
  \BibitemOpen
  \bibfield  {author} {\bibinfo {author} {\bibfnamefont {A.~A.}\ \bibnamefont {Zyuzin}}, \bibinfo {author} {\bibfnamefont {Y.}~\bibnamefont {Wu}},\ and\ \bibinfo {author} {\bibfnamefont {A.~A.}\ \bibnamefont {Burkov}},\ }\bibfield  {title} {\bibinfo {title} {Weyl semimetal with broken time-reversal and inversion symmetries: Relation between quasiparticle lifetime and exceptional points},\ }\href {https://doi.org/10.1103/PhysRevB.97.041203} {\bibfield  {journal} {\bibinfo  {journal} {Physical Review B}\ }\textbf {\bibinfo {volume} {97}},\ \bibinfo {pages} {041203} (\bibinfo {year} {2018})}\BibitemShut {NoStop}%
\bibitem [{\citenamefont {Crippa}\ \emph {et~al.}(2023)\citenamefont {Crippa}, \citenamefont {Sangiovanni},\ and\ \citenamefont {Budich}}]{Crippa2023PRL}%
  \BibitemOpen
  \bibfield  {author} {\bibinfo {author} {\bibfnamefont {L.}~\bibnamefont {Crippa}}, \bibinfo {author} {\bibfnamefont {G.}~\bibnamefont {Sangiovanni}},\ and\ \bibinfo {author} {\bibfnamefont {J.~C.}\ \bibnamefont {Budich}},\ }\bibfield  {title} {\bibinfo {title} {Spontaneous formation of exceptional points at the onset of magnetism},\ }\href {https://doi.org/10.1103/PhysRevLett.130.186403} {\bibfield  {journal} {\bibinfo  {journal} {Phys. Rev. Lett.}\ }\textbf {\bibinfo {volume} {130}},\ \bibinfo {pages} {186403} (\bibinfo {year} {2023})}\BibitemShut {NoStop}%
\bibitem [{\citenamefont {Kozii}\ and\ \citenamefont {Fu}(2024)}]{Kozii_2024}%
  \BibitemOpen
  \bibfield  {author} {\bibinfo {author} {\bibfnamefont {V.}~\bibnamefont {Kozii}}\ and\ \bibinfo {author} {\bibfnamefont {L.}~\bibnamefont {Fu}},\ }\bibfield  {title} {\bibinfo {title} {Non-hermitian topological theory of finite-lifetime quasiparticles: Prediction of bulk fermi arc due to exceptional point},\ }\href {https://doi.org/10.1103/PhysRevB.109.235139} {\bibfield  {journal} {\bibinfo  {journal} {Phys. Rev. B}\ }\textbf {\bibinfo {volume} {109}},\ \bibinfo {pages} {235139} (\bibinfo {year} {2024})}\BibitemShut {NoStop}%
\bibitem [{\citenamefont {Helbig}\ \emph {et~al.}(2020)\citenamefont {Helbig}, \citenamefont {Hofmann}, \citenamefont {Imhof}, \citenamefont {Abdelghany}, \citenamefont {Kiessling}, \citenamefont {Molenkamp}, \citenamefont {Lee}, \citenamefont {Szameit}, \citenamefont {Greiter},\ and\ \citenamefont {Thomale}}]{Helbig2020NatPhys}%
  \BibitemOpen
  \bibfield  {author} {\bibinfo {author} {\bibfnamefont {T.}~\bibnamefont {Helbig}}, \bibinfo {author} {\bibfnamefont {T.}~\bibnamefont {Hofmann}}, \bibinfo {author} {\bibfnamefont {S.}~\bibnamefont {Imhof}}, \bibinfo {author} {\bibfnamefont {M.}~\bibnamefont {Abdelghany}}, \bibinfo {author} {\bibfnamefont {T.}~\bibnamefont {Kiessling}}, \bibinfo {author} {\bibfnamefont {L.~W.}\ \bibnamefont {Molenkamp}}, \bibinfo {author} {\bibfnamefont {C.~H.}\ \bibnamefont {Lee}}, \bibinfo {author} {\bibfnamefont {A.}~\bibnamefont {Szameit}}, \bibinfo {author} {\bibfnamefont {M.}~\bibnamefont {Greiter}},\ and\ \bibinfo {author} {\bibfnamefont {R.}~\bibnamefont {Thomale}},\ }\bibfield  {title} {\bibinfo {title} {Generalized bulk–boundary correspondence in non-hermitian topolectrical circuits},\ }\href {https://doi.org/10.1038/s41567-020-0922-9} {\bibfield  {journal} {\bibinfo  {journal} {Nat. Phys.}\ }\textbf {\bibinfo {volume} {16}},\ \bibinfo {pages} {747} (\bibinfo {year} {2020})}\BibitemShut {NoStop}%
\bibitem [{\citenamefont {Weidemann}\ \emph {et~al.}(2020)\citenamefont {Weidemann}, \citenamefont {Kremer}, \citenamefont {Longhi},\ and\ \citenamefont {Szameit}}]{Weidemann2020Science}%
  \BibitemOpen
  \bibfield  {author} {\bibinfo {author} {\bibfnamefont {S.}~\bibnamefont {Weidemann}}, \bibinfo {author} {\bibfnamefont {M.}~\bibnamefont {Kremer}}, \bibinfo {author} {\bibfnamefont {S.}~\bibnamefont {Longhi}},\ and\ \bibinfo {author} {\bibfnamefont {A.}~\bibnamefont {Szameit}},\ }\bibfield  {title} {\bibinfo {title} {Topological funneling of light},\ }\href {https://doi.org/10.1126/science.aaz8727} {\bibfield  {journal} {\bibinfo  {journal} {Science}\ }\textbf {\bibinfo {volume} {368}},\ \bibinfo {pages} {311} (\bibinfo {year} {2020})}\BibitemShut {NoStop}%
\bibitem [{\citenamefont {Ghatak}\ \emph {et~al.}(2020)\citenamefont {Ghatak}, \citenamefont {Brandenbourger}, \citenamefont {van Wezel},\ and\ \citenamefont {Coulais}}]{ghatak2020}%
  \BibitemOpen
  \bibfield  {author} {\bibinfo {author} {\bibfnamefont {A.}~\bibnamefont {Ghatak}}, \bibinfo {author} {\bibfnamefont {M.}~\bibnamefont {Brandenbourger}}, \bibinfo {author} {\bibfnamefont {J.}~\bibnamefont {van Wezel}},\ and\ \bibinfo {author} {\bibfnamefont {C.}~\bibnamefont {Coulais}},\ }\bibfield  {title} {\bibinfo {title} {Observation of non-hermitian topology and its bulk–edge correspondence in an active mechanical metamaterial},\ }\href {https://doi.org/10.1073/pnas.2010580117} {\bibfield  {journal} {\bibinfo  {journal} {Proceedings of the National Academy of Sciences}\ }\textbf {\bibinfo {volume} {117}},\ \bibinfo {pages} {29561} (\bibinfo {year} {2020})},\ \Eprint {https://arxiv.org/abs/https://www.pnas.org/doi/pdf/10.1073/pnas.2010580117} {https://www.pnas.org/doi/pdf/10.1073/pnas.2010580117} \BibitemShut {NoStop}%
\bibitem [{\citenamefont {Gong}\ \emph {et~al.}(2018)\citenamefont {Gong}, \citenamefont {Ashida}, \citenamefont {Kawabata}, \citenamefont {Takasan}, \citenamefont {Higashikawa},\ and\ \citenamefont {Ueda}}]{Gong2018PRX}%
  \BibitemOpen
  \bibfield  {author} {\bibinfo {author} {\bibfnamefont {Z.}~\bibnamefont {Gong}}, \bibinfo {author} {\bibfnamefont {Y.}~\bibnamefont {Ashida}}, \bibinfo {author} {\bibfnamefont {K.}~\bibnamefont {Kawabata}}, \bibinfo {author} {\bibfnamefont {K.}~\bibnamefont {Takasan}}, \bibinfo {author} {\bibfnamefont {S.}~\bibnamefont {Higashikawa}},\ and\ \bibinfo {author} {\bibfnamefont {M.}~\bibnamefont {Ueda}},\ }\bibfield  {title} {\bibinfo {title} {Topological phases of non-hermitian systems},\ }\href {https://doi.org/10.1103/PhysRevX.8.031079} {\bibfield  {journal} {\bibinfo  {journal} {Phys. Rev. X}\ }\textbf {\bibinfo {volume} {8}},\ \bibinfo {pages} {031079} (\bibinfo {year} {2018})}\BibitemShut {NoStop}%
\bibitem [{\citenamefont {Kawabata}\ \emph {et~al.}(2019)\citenamefont {Kawabata}, \citenamefont {Shiozaki}, \citenamefont {Ueda},\ and\ \citenamefont {Sato}}]{Kawabata2019PRX}%
  \BibitemOpen
  \bibfield  {author} {\bibinfo {author} {\bibfnamefont {K.}~\bibnamefont {Kawabata}}, \bibinfo {author} {\bibfnamefont {K.}~\bibnamefont {Shiozaki}}, \bibinfo {author} {\bibfnamefont {M.}~\bibnamefont {Ueda}},\ and\ \bibinfo {author} {\bibfnamefont {M.}~\bibnamefont {Sato}},\ }\bibfield  {title} {\bibinfo {title} {Symmetry and topology in non-hermitian physics},\ }\href {https://doi.org/10.1103/PhysRevX.9.041015} {\bibfield  {journal} {\bibinfo  {journal} {Phys. Rev. X}\ }\textbf {\bibinfo {volume} {9}},\ \bibinfo {pages} {041015} (\bibinfo {year} {2019})}\BibitemShut {NoStop}%
\bibitem [{\citenamefont {Yao}\ and\ \citenamefont {Wang}(2018)}]{YaoWang2018PRL}%
  \BibitemOpen
  \bibfield  {author} {\bibinfo {author} {\bibfnamefont {S.}~\bibnamefont {Yao}}\ and\ \bibinfo {author} {\bibfnamefont {Z.}~\bibnamefont {Wang}},\ }\bibfield  {title} {\bibinfo {title} {Edge states and topological invariants of non-hermitian systems},\ }\href {https://doi.org/10.1103/PhysRevLett.121.086803} {\bibfield  {journal} {\bibinfo  {journal} {Physical Review Letters}\ }\textbf {\bibinfo {volume} {121}},\ \bibinfo {pages} {086803} (\bibinfo {year} {2018})}\BibitemShut {NoStop}%
\bibitem [{\citenamefont {Lee}\ and\ \citenamefont {Thomale}(2019)}]{LeeThomale2019PRB}%
  \BibitemOpen
  \bibfield  {author} {\bibinfo {author} {\bibfnamefont {C.~H.}\ \bibnamefont {Lee}}\ and\ \bibinfo {author} {\bibfnamefont {R.}~\bibnamefont {Thomale}},\ }\bibfield  {title} {\bibinfo {title} {Anatomy of skin modes and topology in non-hermitian systems},\ }\href {https://doi.org/10.1103/PhysRevB.99.201103} {\bibfield  {journal} {\bibinfo  {journal} {Physical Review B}\ }\textbf {\bibinfo {volume} {99}},\ \bibinfo {pages} {201103} (\bibinfo {year} {2019})}\BibitemShut {NoStop}%
\bibitem [{\citenamefont {Lin}\ \emph {et~al.}(2023)\citenamefont {Lin}, \citenamefont {Tai}, \citenamefont {Li},\ and\ \citenamefont {Lee}}]{lin_topological_2023}%
  \BibitemOpen
  \bibfield  {author} {\bibinfo {author} {\bibfnamefont {R.}~\bibnamefont {Lin}}, \bibinfo {author} {\bibfnamefont {T.}~\bibnamefont {Tai}}, \bibinfo {author} {\bibfnamefont {L.}~\bibnamefont {Li}},\ and\ \bibinfo {author} {\bibfnamefont {C.~H.}\ \bibnamefont {Lee}},\ }\bibfield  {title} {\bibinfo {title} {Topological non-{Hermitian} skin effect},\ }\href {https://doi.org/10.1007/s11467-023-1309-z} {\bibfield  {journal} {\bibinfo  {journal} {Frontiers of Physics}\ }\textbf {\bibinfo {volume} {18}},\ \bibinfo {pages} {53605} (\bibinfo {year} {2023})}\BibitemShut {NoStop}%
\bibitem [{\citenamefont {Yang}\ \emph {et~al.}(2024)\citenamefont {Yang}, \citenamefont {Li}, \citenamefont {König}, \citenamefont {Rødland}, \citenamefont {Stålhammar},\ and\ \citenamefont {Bergholtz}}]{yang2024rpp}%
  \BibitemOpen
  \bibfield  {author} {\bibinfo {author} {\bibfnamefont {K.}~\bibnamefont {Yang}}, \bibinfo {author} {\bibfnamefont {Z.}~\bibnamefont {Li}}, \bibinfo {author} {\bibfnamefont {J.~L.~K.}\ \bibnamefont {König}}, \bibinfo {author} {\bibfnamefont {L.}~\bibnamefont {Rødland}}, \bibinfo {author} {\bibfnamefont {M.}~\bibnamefont {Stålhammar}},\ and\ \bibinfo {author} {\bibfnamefont {E.~J.}\ \bibnamefont {Bergholtz}},\ }\bibfield  {title} {\bibinfo {title} {Homotopy, {{Symmetry}}, and non-{{Hermitian Band Topology}}},\ }\href {https://doi.org/10.1088/1361-6633/ad4e64} {\bibfield  {journal} {\bibinfo  {journal} {Rep. Prog. Phys.}\ }\textbf {\bibinfo {volume} {87}},\ \bibinfo {pages} {078002} (\bibinfo {year} {2024})}\BibitemShut {NoStop}%
\bibitem [{\citenamefont {Buča}\ and\ \citenamefont {Prosen}(2012)}]{buca_2012}%
  \BibitemOpen
  \bibfield  {author} {\bibinfo {author} {\bibfnamefont {B.}~\bibnamefont {Buča}}\ and\ \bibinfo {author} {\bibfnamefont {T.}~\bibnamefont {Prosen}},\ }\bibfield  {title} {\bibinfo {title} {A note on symmetry reductions of the lindblad equation: transport in constrained open spin chains},\ }\href {https://doi.org/10.1088/1367-2630/14/7/073007} {\bibfield  {journal} {\bibinfo  {journal} {New Journal of Physics}\ }\textbf {\bibinfo {volume} {14}},\ \bibinfo {pages} {073007} (\bibinfo {year} {2012})}\BibitemShut {NoStop}%
\bibitem [{\citenamefont {Altland}\ \emph {et~al.}(2021)\citenamefont {Altland}, \citenamefont {Fleischhauer},\ and\ \citenamefont {Diehl}}]{altland_2021}%
  \BibitemOpen
  \bibfield  {author} {\bibinfo {author} {\bibfnamefont {A.}~\bibnamefont {Altland}}, \bibinfo {author} {\bibfnamefont {M.}~\bibnamefont {Fleischhauer}},\ and\ \bibinfo {author} {\bibfnamefont {S.}~\bibnamefont {Diehl}},\ }\bibfield  {title} {\bibinfo {title} {Symmetry classes of open fermionic quantum matter},\ }\href {https://doi.org/10.1103/PhysRevX.11.021037} {\bibfield  {journal} {\bibinfo  {journal} {Phys. Rev. X}\ }\textbf {\bibinfo {volume} {11}},\ \bibinfo {pages} {021037} (\bibinfo {year} {2021})}\BibitemShut {NoStop}%
\bibitem [{\citenamefont {Sá}\ \emph {et~al.}(2023)\citenamefont {Sá}, \citenamefont {Ribeiro},\ and\ \citenamefont {Prosen}}]{sa_symmetry_2023}%
  \BibitemOpen
  \bibfield  {author} {\bibinfo {author} {\bibfnamefont {L.}~\bibnamefont {Sá}}, \bibinfo {author} {\bibfnamefont {P.}~\bibnamefont {Ribeiro}},\ and\ \bibinfo {author} {\bibfnamefont {T.}~\bibnamefont {Prosen}},\ }\bibfield  {title} {\bibinfo {title} {Symmetry {Classification} of {Many}-{Body} {Lindbladians}: {Tenfold} {Way} and {Beyond}},\ }\href {https://doi.org/10.1103/PhysRevX.13.031019} {\bibfield  {journal} {\bibinfo  {journal} {Physical Review X}\ }\textbf {\bibinfo {volume} {13}},\ \bibinfo {pages} {031019} (\bibinfo {year} {2023})},\ \bibinfo {note} {publisher: American Physical Society}\BibitemShut {NoStop}%
\bibitem [{\citenamefont {Haldane}(1988)}]{haldane1988aqhe}%
  \BibitemOpen
  \bibfield  {author} {\bibinfo {author} {\bibfnamefont {F.~D.~M.}\ \bibnamefont {Haldane}},\ }\bibfield  {title} {\bibinfo {title} {Model for a quantum hall effect without landau levels: Condensed-matter realization of the "parity anomaly"},\ }\href {https://doi.org/10.1103/PhysRevLett.61.2015} {\bibfield  {journal} {\bibinfo  {journal} {Phys. Rev. Lett.}\ }\textbf {\bibinfo {volume} {61}},\ \bibinfo {pages} {2015} (\bibinfo {year} {1988})}\BibitemShut {NoStop}%
\bibitem [{\citenamefont {Chang}\ \emph {et~al.}(2023)\citenamefont {Chang}, \citenamefont {Liu},\ and\ \citenamefont {MacDonald}}]{chang2023colloquium}%
  \BibitemOpen
  \bibfield  {author} {\bibinfo {author} {\bibfnamefont {C.-Z.}\ \bibnamefont {Chang}}, \bibinfo {author} {\bibfnamefont {C.-X.}\ \bibnamefont {Liu}},\ and\ \bibinfo {author} {\bibfnamefont {A.~H.}\ \bibnamefont {MacDonald}},\ }\bibfield  {title} {\bibinfo {title} {Colloquium: Quantum anomalous hall effect},\ }\href {https://doi.org/10.1103/RevModPhys.95.011002} {\bibfield  {journal} {\bibinfo  {journal} {Rev. Mod. Phys.}\ }\textbf {\bibinfo {volume} {95}},\ \bibinfo {pages} {011002} (\bibinfo {year} {2023})}\BibitemShut {NoStop}%
\bibitem [{\citenamefont {Liu}\ and\ \citenamefont {Bergholtz}(2024)}]{Liu2024chern}%
  \BibitemOpen
  \bibfield  {author} {\bibinfo {author} {\bibfnamefont {Z.}~\bibnamefont {Liu}}\ and\ \bibinfo {author} {\bibfnamefont {E.~J.}\ \bibnamefont {Bergholtz}},\ }\bibinfo {title} {Recent developments in fractional chern insulators},\ in\ \href {https://doi.org/10.1016/b978-0-323-90800-9.00136-0} {\emph {\bibinfo {booktitle} {Encyclopedia of Condensed Matter Physics}}}\ (\bibinfo  {publisher} {Elsevier},\ \bibinfo {year} {2024})\ p.\ \bibinfo {pages} {515–538}\BibitemShut {NoStop}%
\bibitem [{\citenamefont {Niemi}\ and\ \citenamefont {Semenoff}(1983)}]{niemi_axial-anomaly-induced_1983}%
  \BibitemOpen
  \bibfield  {author} {\bibinfo {author} {\bibfnamefont {A.~J.}\ \bibnamefont {Niemi}}\ and\ \bibinfo {author} {\bibfnamefont {G.~W.}\ \bibnamefont {Semenoff}},\ }\bibfield  {title} {\bibinfo {title} {Axial-{Anomaly}-{Induced} {Fermion} {Fractionization} and {Effective} {Gauge}-{Theory} {Actions} in {Odd}-{Dimensional} {Space}-{Times}},\ }\href {https://doi.org/10.1103/PhysRevLett.51.2077} {\bibfield  {journal} {\bibinfo  {journal} {Physical Review Letters}\ }\textbf {\bibinfo {volume} {51}},\ \bibinfo {pages} {2077} (\bibinfo {year} {1983})},\ \bibinfo {note} {publisher: American Physical Society}\BibitemShut {NoStop}%
\bibitem [{\citenamefont {Redlich}(1984)}]{redlich_gauge_1984}%
  \BibitemOpen
  \bibfield  {author} {\bibinfo {author} {\bibfnamefont {A.~N.}\ \bibnamefont {Redlich}},\ }\bibfield  {title} {\bibinfo {title} {Gauge {Noninvariance} and {Parity} {Nonconservation} of {Three}-{Dimensional} {Fermions}},\ }\href {https://doi.org/10.1103/PhysRevLett.52.18} {\bibfield  {journal} {\bibinfo  {journal} {Physical Review Letters}\ }\textbf {\bibinfo {volume} {52}},\ \bibinfo {pages} {18} (\bibinfo {year} {1984})},\ \bibinfo {note} {publisher: American Physical Society}\BibitemShut {NoStop}%
\bibitem [{\citenamefont {Alvarez-Gaumé}\ and\ \citenamefont {Witten}(1984)}]{alvarez-gaume_gravitational_1984}%
  \BibitemOpen
  \bibfield  {author} {\bibinfo {author} {\bibfnamefont {L.}~\bibnamefont {Alvarez-Gaumé}}\ and\ \bibinfo {author} {\bibfnamefont {E.}~\bibnamefont {Witten}},\ }\bibfield  {title} {\bibinfo {title} {Gravitational anomalies},\ }\href {https://doi.org/10.1016/0550-3213(84)90066-X} {\bibfield  {journal} {\bibinfo  {journal} {Nuclear Physics B}\ }\textbf {\bibinfo {volume} {234}},\ \bibinfo {pages} {269} (\bibinfo {year} {1984})}\BibitemShut {NoStop}%
\bibitem [{\citenamefont {Arouca}\ \emph {et~al.}(2022)\citenamefont {Arouca}, \citenamefont {Cappelli},\ and\ \citenamefont {Hansson}}]{arouca2022anomalies}%
  \BibitemOpen
  \bibfield  {author} {\bibinfo {author} {\bibfnamefont {R.}~\bibnamefont {Arouca}}, \bibinfo {author} {\bibfnamefont {A.}~\bibnamefont {Cappelli}},\ and\ \bibinfo {author} {\bibfnamefont {T.~H.}\ \bibnamefont {Hansson}},\ }\bibfield  {title} {\bibinfo {title} {{Quantum Field Theory Anomalies in Condensed Matter Physics}},\ }\href {https://doi.org/10.21468/SciPostPhysLectNotes.62} {\bibfield  {journal} {\bibinfo  {journal} {SciPost Phys. Lect. Notes}\ ,\ \bibinfo {pages} {62}} (\bibinfo {year} {2022})}\BibitemShut {NoStop}%
\bibitem [{\citenamefont {Janssen}\ \emph {et~al.}(2012)\citenamefont {Janssen}, \citenamefont {Williams}, \citenamefont {Fletcher}, \citenamefont {Goebel}, \citenamefont {Tzalenchuk}, \citenamefont {Yakimova}, \citenamefont {Lara-Avila}, \citenamefont {Kubatkin},\ and\ \citenamefont {Fal'ko}}]{Janssen2012gaas}%
  \BibitemOpen
  \bibfield  {author} {\bibinfo {author} {\bibfnamefont {T.~J. B.~M.}\ \bibnamefont {Janssen}}, \bibinfo {author} {\bibfnamefont {J.~M.}\ \bibnamefont {Williams}}, \bibinfo {author} {\bibfnamefont {N.~E.}\ \bibnamefont {Fletcher}}, \bibinfo {author} {\bibfnamefont {R.}~\bibnamefont {Goebel}}, \bibinfo {author} {\bibfnamefont {A.}~\bibnamefont {Tzalenchuk}}, \bibinfo {author} {\bibfnamefont {R.}~\bibnamefont {Yakimova}}, \bibinfo {author} {\bibfnamefont {S.}~\bibnamefont {Lara-Avila}}, \bibinfo {author} {\bibfnamefont {S.}~\bibnamefont {Kubatkin}},\ and\ \bibinfo {author} {\bibfnamefont {V.~I.}\ \bibnamefont {Fal'ko}},\ }\bibfield  {title} {\bibinfo {title} {Precision comparison of the quantum hall effect in graphene and gallium arsenide},\ }\href {https://doi.org/10.1088/0026-1394/49/3/294} {\bibfield  {journal} {\bibinfo  {journal} {Metrologia}\ }\textbf {\bibinfo {volume} {49}},\ \bibinfo {pages} {294} (\bibinfo {year} {2012})}\BibitemShut {NoStop}%
\bibitem [{\citenamefont {Matsubara}(1955)}]{matsubara1955new}%
  \BibitemOpen
  \bibfield  {author} {\bibinfo {author} {\bibfnamefont {T.}~\bibnamefont {Matsubara}},\ }\bibfield  {title} {\bibinfo {title} {A new approach to quantum-statistical mechanics},\ }\href@noop {} {\bibfield  {journal} {\bibinfo  {journal} {Progress of theoretical physics}\ }\textbf {\bibinfo {volume} {14}},\ \bibinfo {pages} {351} (\bibinfo {year} {1955})}\BibitemShut {NoStop}%
\bibitem [{\citenamefont {Schwinger}(1960)}]{schwinger1960special}%
  \BibitemOpen
  \bibfield  {author} {\bibinfo {author} {\bibfnamefont {J.}~\bibnamefont {Schwinger}},\ }\bibfield  {title} {\bibinfo {title} {The special canonical group},\ }\href@noop {} {\bibfield  {journal} {\bibinfo  {journal} {Proceedings of the National Academy of Sciences}\ }\textbf {\bibinfo {volume} {46}},\ \bibinfo {pages} {1401} (\bibinfo {year} {1960})}\BibitemShut {NoStop}%
\bibitem [{\citenamefont {Keldysh}(1964)}]{keldysh1964diagram}%
  \BibitemOpen
  \bibfield  {author} {\bibinfo {author} {\bibfnamefont {L.~V.}\ \bibnamefont {Keldysh}},\ }\bibfield  {title} {\bibinfo {title} {Diagram technique for nonequilibrium processes},\ }\href@noop {} {\bibfield  {journal} {\bibinfo  {journal} {Zh. Eksp. Teor. Fiz}\ }\textbf {\bibinfo {volume} {47}},\ \bibinfo {pages} {151} (\bibinfo {year} {1964})}\BibitemShut {NoStop}%
\bibitem [{\citenamefont {Kadanoff}\ and\ \citenamefont {Baym}(1962)}]{kadanoff1962quantum}%
  \BibitemOpen
  \bibfield  {author} {\bibinfo {author} {\bibfnamefont {L.~P.}\ \bibnamefont {Kadanoff}}\ and\ \bibinfo {author} {\bibfnamefont {G.~A.}\ \bibnamefont {Baym}},\ }\href@noop {} {\emph {\bibinfo {title} {Quantum Statistical Mechanics Green's Function Methods in Equilibrium Problems}}}\ (\bibinfo  {publisher} {Benjamin},\ \bibinfo {year} {1962})\BibitemShut {NoStop}%
\bibitem [{\citenamefont {Sieberer}\ \emph {et~al.}(2016)\citenamefont {Sieberer}, \citenamefont {Buchhold},\ and\ \citenamefont {Diehl}}]{sieberer_keldysh_2016}%
  \BibitemOpen
  \bibfield  {author} {\bibinfo {author} {\bibfnamefont {L.~M.}\ \bibnamefont {Sieberer}}, \bibinfo {author} {\bibfnamefont {M.}~\bibnamefont {Buchhold}},\ and\ \bibinfo {author} {\bibfnamefont {S.}~\bibnamefont {Diehl}},\ }\bibfield  {title} {\bibinfo {title} {Keldysh field theory for driven open quantum systems},\ }\href {https://doi.org/10.1088/0034-4885/79/9/096001} {\bibfield  {journal} {\bibinfo  {journal} {Reports on Progress in Physics}\ }\textbf {\bibinfo {volume} {79}},\ \bibinfo {pages} {096001} (\bibinfo {year} {2016})},\ \bibinfo {note} {publisher: IOP Publishing}\BibitemShut {NoStop}%
\bibitem [{\citenamefont {Haehl}\ \emph {et~al.}(2017)\citenamefont {Haehl}, \citenamefont {Loganayagam},\ and\ \citenamefont {Rangamani}}]{haehl_schwinger-keldysh_2017}%
  \BibitemOpen
  \bibfield  {author} {\bibinfo {author} {\bibfnamefont {F.~M.}\ \bibnamefont {Haehl}}, \bibinfo {author} {\bibfnamefont {R.}~\bibnamefont {Loganayagam}},\ and\ \bibinfo {author} {\bibfnamefont {M.}~\bibnamefont {Rangamani}},\ }\bibfield  {title} {\bibinfo {title} {Schwinger-{Keldysh} formalism. {Part} {I}: {BRST} symmetries and superspace},\ }\href {https://doi.org/10.1007/JHEP06(2017)069} {\bibfield  {journal} {\bibinfo  {journal} {Journal of High Energy Physics}\ }\textbf {\bibinfo {volume} {2017}},\ \bibinfo {pages} {69} (\bibinfo {year} {2017})}\BibitemShut {NoStop}%
\bibitem [{\citenamefont {Kamenev}(2023)}]{kamenev}%
  \BibitemOpen
  \bibfield  {author} {\bibinfo {author} {\bibfnamefont {A.}~\bibnamefont {Kamenev}},\ }\href@noop {} {\emph {\bibinfo {title} {Field Theory of Non-Equilibrium Systems}}},\ \bibinfo {edition} {2nd}\ ed.\ (\bibinfo  {publisher} {Cambridge University Press},\ \bibinfo {year} {2023})\BibitemShut {NoStop}%
\bibitem [{\citenamefont {Sieberer}\ \emph {et~al.}(2025)\citenamefont {Sieberer}, \citenamefont {Buchhold}, \citenamefont {Marino},\ and\ \citenamefont {Diehl}}]{sieberer_universality_2025}%
  \BibitemOpen
  \bibfield  {author} {\bibinfo {author} {\bibfnamefont {L.~M.}\ \bibnamefont {Sieberer}}, \bibinfo {author} {\bibfnamefont {M.}~\bibnamefont {Buchhold}}, \bibinfo {author} {\bibfnamefont {J.}~\bibnamefont {Marino}},\ and\ \bibinfo {author} {\bibfnamefont {S.}~\bibnamefont {Diehl}},\ }\bibfield  {title} {\bibinfo {title} {Universality in driven open quantum matter},\ }\href {https://doi.org/10.1103/RevModPhys.97.025004} {\bibfield  {journal} {\bibinfo  {journal} {Reviews of Modern Physics}\ }\textbf {\bibinfo {volume} {97}},\ \bibinfo {pages} {025004} (\bibinfo {year} {2025})},\ \bibinfo {note} {publisher: American Physical Society}\BibitemShut {NoStop}%
\bibitem [{\citenamefont {Coleman}(2015)}]{coleman}%
  \BibitemOpen
  \bibfield  {author} {\bibinfo {author} {\bibfnamefont {P.}~\bibnamefont {Coleman}},\ }\href@noop {} {\emph {\bibinfo {title} {Introduction to Many-Body Physics}}}\ (\bibinfo  {publisher} {Cambridge University Press},\ \bibinfo {year} {2015})\BibitemShut {NoStop}%
\bibitem [{\citenamefont {Kohri}\ and\ \citenamefont {Yamada}(2002)}]{kohri2002polarization}%
  \BibitemOpen
  \bibfield  {author} {\bibinfo {author} {\bibfnamefont {K.}~\bibnamefont {Kohri}}\ and\ \bibinfo {author} {\bibfnamefont {S.}~\bibnamefont {Yamada}},\ }\bibfield  {title} {\bibinfo {title} {Polarization tensors in strong magnetic fields},\ }\href {https://doi.org/10.1103/PhysRevD.65.043006} {\bibfield  {journal} {\bibinfo  {journal} {Phys. Rev. D}\ }\textbf {\bibinfo {volume} {65}},\ \bibinfo {pages} {043006} (\bibinfo {year} {2002})}\BibitemShut {NoStop}%
\bibitem [{\citenamefont {Gorbar}\ \emph {et~al.}(2002)\citenamefont {Gorbar}, \citenamefont {Gusynin}, \citenamefont {Miransky},\ and\ \citenamefont {Shovkovy}}]{gorbar2002magnetic}%
  \BibitemOpen
  \bibfield  {author} {\bibinfo {author} {\bibfnamefont {E.~V.}\ \bibnamefont {Gorbar}}, \bibinfo {author} {\bibfnamefont {V.~P.}\ \bibnamefont {Gusynin}}, \bibinfo {author} {\bibfnamefont {V.~A.}\ \bibnamefont {Miransky}},\ and\ \bibinfo {author} {\bibfnamefont {I.~A.}\ \bibnamefont {Shovkovy}},\ }\bibfield  {title} {\bibinfo {title} {Magnetic field driven metal-insulator phase transition in planar systems},\ }\href {https://doi.org/10.1103/PhysRevB.66.045108} {\bibfield  {journal} {\bibinfo  {journal} {Phys. Rev. B}\ }\textbf {\bibinfo {volume} {66}},\ \bibinfo {pages} {045108} (\bibinfo {year} {2002})}\BibitemShut {NoStop}%
\bibitem [{\citenamefont {Fr\"a\ss{}dorf}(2018)}]{frassdorf2018abelian}%
  \BibitemOpen
  \bibfield  {author} {\bibinfo {author} {\bibfnamefont {C.}~\bibnamefont {Fr\"a\ss{}dorf}},\ }\bibfield  {title} {\bibinfo {title} {Abelian chern-simons theory for the fractional quantum hall effect in graphene},\ }\href {https://doi.org/10.1103/PhysRevB.97.115123} {\bibfield  {journal} {\bibinfo  {journal} {Phys. Rev. B}\ }\textbf {\bibinfo {volume} {97}},\ \bibinfo {pages} {115123} (\bibinfo {year} {2018})}\BibitemShut {NoStop}%
\bibitem [{\citenamefont {Mulligan}\ and\ \citenamefont {Burnell}(2013)}]{MulliganBurnell2013}%
  \BibitemOpen
  \bibfield  {author} {\bibinfo {author} {\bibfnamefont {M.}~\bibnamefont {Mulligan}}\ and\ \bibinfo {author} {\bibfnamefont {F.~J.}\ \bibnamefont {Burnell}},\ }\bibfield  {title} {\bibinfo {title} {Topological insulators avoid the parity anomaly},\ }\href {https://doi.org/10.1103/PhysRevB.88.085104} {\bibfield  {journal} {\bibinfo  {journal} {Phys. Rev. B}\ }\textbf {\bibinfo {volume} {88}},\ \bibinfo {pages} {085104} (\bibinfo {year} {2013})}\BibitemShut {NoStop}%
\bibitem [{\citenamefont {Callan}\ and\ \citenamefont {Harvey}(1985)}]{CALLAN1985427}%
  \BibitemOpen
  \bibfield  {author} {\bibinfo {author} {\bibfnamefont {C.}~\bibnamefont {Callan}}\ and\ \bibinfo {author} {\bibfnamefont {J.}~\bibnamefont {Harvey}},\ }\bibfield  {title} {\bibinfo {title} {Anomalies and fermion zero modes on strings and domain walls},\ }\href {https://doi.org/https://doi.org/10.1016/0550-3213(85)90489-4} {\bibfield  {journal} {\bibinfo  {journal} {Nuclear Physics B}\ }\textbf {\bibinfo {volume} {250}},\ \bibinfo {pages} {427} (\bibinfo {year} {1985})}\BibitemShut {NoStop}%
\bibitem [{\citenamefont {Mogi}\ \emph {et~al.}(2022)\citenamefont {Mogi}, \citenamefont {Okamura}, \citenamefont {Kawamura}, \citenamefont {Yoshimi}, \citenamefont {Yasuda}, \citenamefont {Tsukazaki}, \citenamefont {Takahashi}, \citenamefont {Morimoto}, \citenamefont {Nagaosa}, \citenamefont {Kawasaki}, \citenamefont {Takahashi},\ and\ \citenamefont {Tokura}}]{mogi_experimental_2022}%
  \BibitemOpen
  \bibfield  {author} {\bibinfo {author} {\bibfnamefont {M.}~\bibnamefont {Mogi}}, \bibinfo {author} {\bibfnamefont {Y.}~\bibnamefont {Okamura}}, \bibinfo {author} {\bibfnamefont {M.}~\bibnamefont {Kawamura}}, \bibinfo {author} {\bibfnamefont {R.}~\bibnamefont {Yoshimi}}, \bibinfo {author} {\bibfnamefont {K.}~\bibnamefont {Yasuda}}, \bibinfo {author} {\bibfnamefont {A.}~\bibnamefont {Tsukazaki}}, \bibinfo {author} {\bibfnamefont {K.~S.}\ \bibnamefont {Takahashi}}, \bibinfo {author} {\bibfnamefont {T.}~\bibnamefont {Morimoto}}, \bibinfo {author} {\bibfnamefont {N.}~\bibnamefont {Nagaosa}}, \bibinfo {author} {\bibfnamefont {M.}~\bibnamefont {Kawasaki}}, \bibinfo {author} {\bibfnamefont {Y.}~\bibnamefont {Takahashi}},\ and\ \bibinfo {author} {\bibfnamefont {Y.}~\bibnamefont {Tokura}},\ }\bibfield  {title} {\bibinfo {title} {Experimental signature of the parity anomaly in a semi-magnetic topological insulator},\ }\href {https://doi.org/10.1038/s41567-021-01490-y} {\bibfield  {journal} {\bibinfo  {journal} {Nature
  Physics}\ }\textbf {\bibinfo {volume} {18}},\ \bibinfo {pages} {390} (\bibinfo {year} {2022})}\BibitemShut {NoStop}%
\bibitem [{\citenamefont {Sokhotskii}(1873)}]{sokhotskii1873definite}%
  \BibitemOpen
  \bibfield  {author} {\bibinfo {author} {\bibfnamefont {Y.}~\bibnamefont {Sokhotskii}},\ }\bibfield  {title} {\bibinfo {title} {On definite integrals and functions used in series expansions},\ }\href@noop {} {\bibfield  {journal} {\bibinfo  {journal} {published by M. Stasyulevich, St. Petersburg}\ } (\bibinfo {year} {1873})}\BibitemShut {NoStop}%
\bibitem [{\citenamefont {Plemelj}(1964)}]{plemelj1964problems}%
  \BibitemOpen
  \bibfield  {author} {\bibinfo {author} {\bibfnamefont {J.}~\bibnamefont {Plemelj}},\ }\href@noop {} {\bibinfo {title} {Problems in the sense of {R}iemann and {K}lein}} (\bibinfo {year} {1964})\BibitemShut {NoStop}%
\bibitem [{\citenamefont {Zee}(2010)}]{zee2010quantum}%
  \BibitemOpen
  \bibfield  {author} {\bibinfo {author} {\bibfnamefont {A.}~\bibnamefont {Zee}},\ }\href@noop {} {\emph {\bibinfo {title} {Quantum field theory in a nutshell}}},\ Vol.~\bibinfo {volume} {7}\ (\bibinfo  {publisher} {Princeton university press},\ \bibinfo {year} {2010})\BibitemShut {NoStop}%
\end{thebibliography}%
\appendix
\begin{widetext}
\section{Derivation of the Lindblad term}\label{sec:LindbladApp}
Here, we derive the non-equilibrium action \eqref{SB'} using the jump operators \eqref{jump} and the definitions \eqref{gpm} of the coefficients $g_+$ and $g_-$. We also show how the quadratic form in eq. \eqref{eq:G0} is modified by the jump operators.

Let us examine the jump operator terms in \eqref{Lterm}. From eq. \eqref{jump}, the operator $L_{\alpha}=\sqrt{g_{L}}\left(i(\sigma_z\psi)_{\alpha}+g_B\phi\psi_{\alpha}\right)$ has $L^{\dagger}_{\alpha}=\sqrt{g_{L}}\left(-i\alpha\psi^{\dagger}_{\alpha}+g_B\phi^{\dagger}\psi^{\dagger}_{\alpha}\right)$. This gives contributions (sums over $\alpha$ throughout)
\begin{eqnarray}
    L_{\alpha}(\psi_+)L_{\alpha}^{\dagger}(-\psi_-)&=&g_{L}[i\alpha\psi_{\alpha+}+g_B\phi_+\psi_{\alpha_+}][i\alpha\psi^{\dagger}_{\alpha-}-g_B\phi^{\dagger}_-\psi^{\dagger}_{\alpha_-}]\nonumber\\
 &=&-g_L\psi_{\alpha_+}\psi_{\alpha_-}^{\dagger}+ig_{L}g_B\alpha(\phi_+\psi_{\alpha+}\psi^{\dagger}_{\alpha-}-\phi_-^{\dagger}\psi_{\alpha+}\psi_{\alpha-}^{\dagger})+...\nonumber\\
    &=&-g_L\psi_{+}\psi_{-}^{\dagger}-ig_{L}g_B(\phi_+-\phi_-)\overline{\psi}_-\psi_++...\nonumber\\
    &=&-g_L\psi_{+}\psi_{-}^{\dagger}-\sqrt{2}ig_{L}g_B\phi_q\overline{\psi}_-\psi_++...,
\end{eqnarray}
since $\alpha^2=1$. Above, we have specialized to real boson fields and assumed that fermions on different contours anticommute, in addition to using that $\alpha\psi^{\dagger}_{\alpha}\psi_{\alpha}=(\sigma_z\psi^{\dagger})_{\alpha}\psi_{\alpha}=\overline{\psi}\psi$, with a sum over $\alpha$. We ignore the part which is quadratic in the bosons, since this term is quadratic in $g_B$, and hence suppressed in the small-$g_B$ regime. 
\newline The other jump operator type similarly gives
\begin{eqnarray}
    L'_{\alpha}(\psi_+)(L')_{\alpha}^{\dagger}(-\psi_-)&=&g_{G}\left(-i\alpha\psi^{\dagger}_{\alpha+}+g_B\phi_+\psi^{\dagger}_{\alpha_+}\right)\left(-i\alpha\psi_{\alpha-}-g_B\phi^{\dagger}_-\psi_{\alpha_-}\right)\nonumber\\
 &=&-g_G\alpha^2\psi_{\alpha+}^{\dagger}\psi_{\alpha_-}-ig_{G}g_B\alpha\left(\phi_+\psi^{\dagger}_{\alpha+}\psi_{\alpha-}-\phi_-^{\dagger}\psi_{\alpha+}^{\dagger}\psi_{\alpha-}\right)+...\nonumber\\
    &=&-g_G\psi_{\alpha+}^{\dagger}\psi_{\alpha_-} -ig_{G}g_B\left(\phi_+-\phi_-\right)\overline{\psi}_+\psi_-+...\nonumber\\
    &=&-g_G\psi_{\alpha+}^{\dagger}\psi_{\alpha_-}-\sqrt{2}ig_{G}g_B\phi_q\overline{\psi}_+\psi_-+...
\end{eqnarray}
Therefore, the first-order terms in $\phi$ that mix the chiralities are
\begin{eqnarray}
    -\sqrt{2}ig_B\phi_q(g_G\overline{\psi}_+\psi_-+g_L\overline{\psi}_-\psi_+).
\end{eqnarray}
After the Keldysh rotation \eqref{rotate} and the use of \eqref{gpm}, these terms give 
\begin{eqnarray}
    -\sqrt{2}ig_B\phi_q\big[g_+(\overline{\psi}_{cl}\psi_{cl}-\overline{\psi}_q\psi_q)+g_-(\overline{\psi}_{q}\psi_{cl}-\overline{\psi}_{cl}\psi_q)\big]    
\end{eqnarray}
as advertised in \eqref{SB'}.

The terms without any boson fields above are also relevant, since they affect the fermionic Green's function in \eqref{eq:G0}. These terms are
\begin{align}
    -g_{L}\psi_{\alpha+}\psi^{\dagger}_{\alpha-}-g_G\psi_{\alpha+}^{\dagger}\psi_{\alpha-}&=-(g_{G}\psi_{\alpha+}^{\dagger}\psi_{\alpha-}-g_L\psi^{\dagger}_{\alpha-}\psi_{\alpha+}),
    \end{align}
and because \eqref{gpm} implies that 
   $g_{G}= g_++g_-$ and $g_L= g_+-g_-$
the above can be rewritten as 
\begin{eqnarray}\label{quadr1}
    &-&(g_{+}+g_-)\psi_{\alpha+}^{\dagger}\psi_{\alpha-}+(g_+-g_-)\psi^{\dagger}_{\alpha-}\psi_{\alpha+}=-g_+(\psi^{\dagger}_q\psi_{cl}-\psi_{cl}^{\dagger}\psi_q)-g_-(\psi^{\dagger}_{cl}\psi_{cl}-\psi^{\dagger}_q\psi_q).
\end{eqnarray}
We remark that in the situation where gain and loss are equal, the second term vanishes. Including the terms $L^{\dagger}_+L_+,L^{\dagger}_-L_-$, etc. in \eqref{Lterm}, we get the following terms quadratic in the fields: 
\begin{align}
    L^{\dagger}_{\alpha}(\psi_+^*,\psi_+)L_{\alpha}(\psi_+^*,\psi_+)&=g_{L}\left(-i\alpha\psi^{\dagger}_{\alpha+}+g_B\phi^{\dagger}\psi^{\dagger}_{\alpha+}\right)\times\left(i\alpha\psi_{\alpha+}+g_B\phi\psi_{\alpha+}\right)=g_L\psi_{+}^{\dagger}\psi_{+}+...,
\end{align}
\begin{align}
    L^{\dagger}_{\alpha}(-\psi_-^*,-\psi_-)L_{\alpha}(-\psi_-^*,-\psi_-)&=g_{L}\left(-i\alpha\psi^{\dagger}_{\alpha-}+g_B\phi^{\dagger}\psi^{\dagger}_{\alpha-}\right)\left(i\alpha\psi_{\alpha-}+g_B\phi\psi_{\alpha-}\right)=g_L\psi_{-}^{\dagger}\psi_{-}+...,
\end{align}
\begin{align}
     L'^{\dagger}_{\alpha}(\psi_+^*,\psi_+)L'_{\alpha}(\psi_+^*,\psi_+)&=g_G\left(i\alpha\psi_{\alpha+}+g_B\phi\psi_{\alpha+}\right)\left(-i\alpha\psi^{\dagger}_{\alpha+}+g_B\phi\psi^{\dagger}_{\alpha+}\right)=g_G\psi_+\psi^{\dagger}_++...,
\end{align}
\begin{align}
     L'^{\dagger}_{\alpha}(-\psi_-^*,-\psi_-)L'_{\alpha}(-\psi_-^*,-\psi_-)&=g_G\left(i\alpha\psi_{\alpha-}+g_B\phi\psi_{\alpha-}\right)
     \left(-i\alpha\psi^{\dagger}_{\alpha-}+g_B\phi\psi^{\dagger}_{\alpha-}\right)=g_G\psi_-\psi^{\dagger}_-+...,
\end{align}
and adding up the four equations above gives a total contribution of 
\begin{eqnarray}\label{quadr2}
    g_-(\psi_{cl}^{\dagger}\psi_{cl}+\psi^{\dagger}_q\psi_q).
\end{eqnarray} 
All in all, \eqref{quadr1} and \eqref{quadr2} result in the extra quadratic terms
\begin{equation}
    2g_-\psi^{\dagger}_q\psi_q+g_+(\psi^{\dagger}_{cl}\psi_q-\psi^{\dagger}_q\psi_{cl})
\end{equation}
from the non-equilibrium effects. Together with the extra factor of $-i$ from \eqref{laction2}, this implies the modified quadratic form in eq. \eqref{eq:G0}, finishing our derivation.

\section{The self-energy}\label{sec:selfenergyApp}
In this section, we derive the self-energy expression (equations \eqref{sigmai} and \eqref{sigmar}) in more detail than in the main text. The results are organized by component, with $\Sigma^{A}$ in section \ref{sec:SA}, $\Sigma^R$ in section \ref{sec:SR} and $\Sigma^K$ in section \ref{sec:SK}. 
\subsection{Advanced self-energy component}\label{sec:SA}
The only diagrams that end up mixing chiralities (i.e. having $\sigma_z$ matrix prefactors) for $\Sigma^A$ are the ones in Fig. \ref{fig:SigmaA}. 
The diagrams are computed below.
\subsubsection{The first two diagrams in Fig. \ref{fig:SigmaA}}\label{copysec}
In terms of the dissipation-modified Green's functions in \eqref{eq:G0}, the integral to be computed is
\begin{eqnarray}
     &ig_Sg_Bg_-\sigma_z\int \frac{d\omega}{2\pi}\frac{d^2k}{(2\pi)^2} D^{R/A}(\omega,k)G^{A}(\nu-\omega,p-k)\nonumber\\&=ig_Sg_Bg_-\sigma_z\int \frac{d\omega}{2\pi}\frac{d^2k}{(2\pi)^2} \frac{1}{2}\frac{1}{\left(\omega\pm i \eta\right)^2-(c_s^2 k^2+m^2)}\frac{1}{(\nu-\omega- ig_+)^2-v_F^2(p-k)^2}\bigg((\nu-\omega- ig_+)I+v_F(\vec{p}-\vec{k})\cdot \vec{\sigma}\bigg).
\end{eqnarray}
Only the identity part gives us the desired $\sigma_z$ prefactor. Keeping only that, and setting $p=0$, we have
\begin{eqnarray}
ig_Sg_Bg_-\sigma_z\int \frac{d\omega}{2\pi}\frac{d^2k}{(2\pi)^2} \frac{1}{2}\frac{1}{\left(\omega\pm i \eta\right)^2-(c_s^2 k^2+m^2)}\frac{1}{(\nu-\omega- ig_+)^2-v_F^2k^2}(\nu-\omega- ig_+).
\end{eqnarray}
We let $\Delta:=\sqrt{c_s^2k^2+m^2}$ and write
\begin{align}
    \frac{1}{(\omega\pm i\eta)^2-\Delta^2}&=\bigg(\frac{1}{\omega\pm i\eta-\Delta}-\frac{1}{\omega \pm i\eta + \Delta}\bigg)\frac{1}{2\Delta}\nonumber\\
    &=\mp \frac{i\pi}{2\Delta} \big(\delta(\omega-\Delta)-\delta(\omega+\Delta)\big)+\mathcal{P}_{\omega=\Delta}\bigg(\frac{1}{2\Delta(\omega-\Delta)}\bigg)-\mathcal{P}_{\omega=-\Delta}\bigg(\frac{1}{2\Delta(\omega+\Delta)}\bigg),
\end{align}
which is valid as an identity for integrands (where the principal value is taken of the integral as a whole). This follows from the Sokhotski-Plemelj theorem.\cite{sokhotskii1873definite,plemelj1964problems} Because we add the diagrams as $D^RG^{A}+D^{A}G^{A}$, and the delta terms only differ in the overall sign, their contributions cancel. We therefore do not compute them. For the principal value terms, we interpret the p.v. at $\pm\Delta$ as
\begin{align}    p.v.=\mathcal{P}\int_a^b=\lim_{\epsilon\rightarrow 0+}\bigg(\int_{a}^{\pm \Delta-\epsilon}+\int_{\pm \Delta+\epsilon}^b\bigg)
\end{align}
The resulting integrals have integrands going to $0$ as $|\omega|\rightarrow 0$, so we may add a contour at infinity without changing the integral, and close it with an integral $I_{\epsilon}$ along a small semicircle above the pole, giving
\begin{eqnarray}
    p.v.+I_{\epsilon}+I_{\infty}=2\pi i\sum \mathrm{Res}(\mathrm{upper \;half-plane})
\end{eqnarray}
We have $I_{\infty}=0$. The epsilon arc integral gives $-i\pi$ times the residue at $\pm\Delta$, so the principal value becomes
\begin{eqnarray}
    p.v.=2\pi i\sum \mathrm{Res}(\mathrm{upper \;half-plane})+i\pi \mathrm{Res}(\pm\Delta).
\end{eqnarray}
Since
\begin{eqnarray}
    \frac{1}{(\nu-\omega- ig_+)^2-v_F^2k^2}=\bigg(\frac{1}{\nu-\omega- ig_+-v_Fk}-\frac{1}{\nu-\omega- ig_++v_Fk}\bigg)\frac{1}{2v_Fk},
\end{eqnarray}
this part gives no residues in the upper half-plane for $g_+>0$. Therefore, the principal value term at $\Delta$ gives (excluding prefactors of the integral)
\begin{align}\label{residuecopy}
    p.v. = i\pi\mathrm{Res}(\Delta)&=
    -\frac{1}{8}g_Sg_Bg_-\sigma_z\int \frac{d^2k}{(2\pi)^2} \frac{1}{\Delta}\bigg(
    \frac{\nu-\Delta- ig_+}{(\nu-\Delta- ig_+)^2-v_F^2k^2}-\frac{\nu+\Delta- ig_+}{(\nu+\Delta- ig_+)^2-v_F^2k^2}\bigg)\nonumber\\
    &=- 
    \frac{1}{16\pi}g_Sg_Bg_-\sigma_z\int dk\frac{k}{\Delta}\bigg(
    \frac{\nu-\Delta- ig_+}{(\nu-\Delta- ig_+)^2-v_F^2k^2}-\frac{\nu+\Delta- ig_+}{(\nu+\Delta- ig_+)^2-v_F^2k^2}\bigg),
\end{align} 
from angular symmetry of the integrand. We series expand the above in $\nu$ to first order, and integrate it from $k=0$ to a UV cutoff $\Lambda$. 
We then add the terms, and use that the Dirac delta terms from the integration cancel between the $G^{A}D^{R/A}$ diagrams when the diagrams are added. The result is complicated, but in the limit of a vanishing bosonic mass parameter $m=0$ and a flat bosonic dispersion $c\rightarrow 0^+$ (or when expanding in $c/v_F$ to linear order), it simplifies to 
 \begin{eqnarray}\label{GADAres}
     -\frac{1}{16\pi}g_Sg_Bg_-\sigma_z\bigg[-\frac{2\Lambda^2}{g_+^2+v_F^2\Lambda^2}+\frac{1}{v_F^2}\log\bigg(1+\frac{v_F^2\Lambda^2}{g_+^2}\bigg)+\frac{2i\Lambda^2(g_+^2-v_F^2\Lambda^2)}{g_+(g_+^2+v_F^2\Lambda^2)^2}\nu\bigg]+\mathcal{O}(\nu^2).
 \end{eqnarray}
 \subsubsection{The third diagram in Fig. \ref{fig:SigmaA}}
We now compute the diagram
\begin{eqnarray}
    &ig_Sg_Bg_+\sigma_z\int \frac{d\omega}{2\pi}\frac{d^2k}{(2\pi)^2} D^{A}(\nu-\omega,p-k)G^{K}(\omega,k)\nonumber\\&=ig_Sg_+\sigma_z\int \frac{d\omega}{2\pi}\frac{d^2k}{(2\pi)^2} \frac{1}{2}\frac{1}{\left(\nu-\omega- i \eta\right)^2-(c_s^2 (p-k)^2+m^2)}\frac{2ig_-}{(\omega^2+v_F^2k^2+g_+^2)^2-4v_F^2\omega^2k^2}\big[(\omega^2+v_F^2k^2+g_+^2)I+2v_F\omega \vec{k}\cdot \vec{\sigma}\big]
\end{eqnarray}
We set $p=0,\Delta=\sqrt{c_s^2k^2+m^2},$ and keep the identity part only, giving
\begin{align}
    ig_Sg_Bg_+(ig_-)\sigma_z\int \frac{d\omega}{2\pi}\frac{d^2k}{(2\pi)^2} \frac{1}{\left(\nu-\omega- i \eta\right)^2-\Delta^2}\frac{1}{(\omega^2+v_F^2k^2+g_+^2)^2-4v_F^2\omega^2k^2}(\omega^2+v_F^2k^2+g_+^2)
\end{align}
The poles of the second factor are at $\omega=\pm vk\pm ig_+$ (where the signs are independent, giving four poles in total). Thus, that factor is defined and continuous on the real line, and we may write
\begin{align}\label{sokhotsky}
   \frac{1}{\left(\nu-\omega- i \eta\right)^2-\Delta^2}&=-\frac{1}{2\Delta}\bigg(\frac{1}{-\nu+\omega+ i \eta+\Delta}-\frac{1}{-\nu+\omega+ i \eta-\Delta}\bigg)\nonumber\\
   &=\frac{i\pi}{2\Delta}\delta(\omega-\nu+\Delta)-\frac{i\pi}{2\Delta}\delta(\omega-\nu-\Delta)+\mathcal{P}\bigg(\frac{1}{2\Delta(\omega-\nu-\Delta)}\bigg)-\mathcal{P}\bigg(\frac{1}{2\Delta(\omega-\nu+\Delta)}\bigg).
\end{align}
This time, there is only one diagram, and the delta terms will therefore not cancel between diagrams. The delta terms, with signs included, each lead to an integrand 
\begin{eqnarray}\label{deltasokhot}
    \pm\frac{i\pi}{2\Delta}\bigg(\frac{(\nu\mp\Delta)^2+v_F^2k^2+g_+^2}{(\nu\mp\Delta)^2+v_F^2k^2+g_+^2)^2-4v_F^2k^2(\nu\mp\Delta)^2}\bigg).
\end{eqnarray}
For the principal values, we stop $\epsilon$ away from the singularity on either side as before. Our function decays fast enough to allow us to close the contour at infinity without contribution, and add an epsilon arc over the singularity (either at $\Delta$ or at $-\Delta$, but not both, since each p.v. only contains one of these singularities). As before, we obtain 
\begin{eqnarray}
    p.v.=2\pi i\sum \mathrm{Res}(\mathrm{upper \;half \;plane})+i\pi \mathrm{Res}(\omega=\nu\pm\Delta).
\end{eqnarray}
The remaining factors in each principal value integral (with principal value around the singularity $\omega=\nu\pm \Delta$) are
\begin{eqnarray}
    \frac{1}{2\Delta(\omega-\nu\mp\Delta)}(\omega^2+v_F^2k^2+g_+^2)\frac{1}{\omega-v_Fk-ig_+}\frac{1}{\omega+v_Fk-ig_+}\frac{1}{\omega-v_Fk+ig_+}\frac{1}{\omega+v_Fk+ig_+},
\end{eqnarray}
so the sum of residues in the upper half-plane (real line excluded) becomes
\begin{align}\label{upperres}
    2\pi i\frac{1}{8} \frac{1}{2\Delta}\bigg[\frac{2v_Fk(v_Fk+ig_+)}{v_Fkig_+(v_Fk+ig_+)}\frac{1}{v_Fk+ig_+-\nu\mp \Delta}+\frac{2v_Fk(v_Fk-ig_+)}{v_Fkig_+(v_Fk-ig_+)}\frac{1}{-v_Fk+ig_+-\nu\mp \Delta}\bigg]\nonumber\\=\frac{\pi}{2g_+\Delta}\frac{ig_+-\nu\mp\Delta}{(ig_+-\nu\mp\Delta)^2-v_F^2k^2}.
\end{align}
The residue at the real-line singularity leads to
\begin{eqnarray}\label{realsing}
    \frac{i\pi}{2\Delta}\frac{(\nu\pm\Delta)^2+v_F^2k^2+g_+^2}{((\nu\pm\Delta)^2+v_F^2k^2+g_+^2)^2-4v_F^2k^2(\nu\pm\Delta)^2
    }
\end{eqnarray}
at $\omega=\nu\pm \Delta$. With the sign $\pm$ of the corresponding p.v. integral in \eqref{sokhotsky}, these terms inherit the $\pm$ sign, which makes them cancel the terms in \eqref{deltasokhot} pairwise. All that remains is the terms from \eqref{upperres}, which inherit a sign $\pm$, meaning that our final integral is
\begin{align}
    &ig_Sg_Bg_+(ig_-)\sigma_z\frac{\pi}{2g_+}\int \frac{1}{2\pi}\frac{d^2k}{(2\pi)^2} \frac{1}{\Delta}\bigg[\frac{ig_+-\nu-\Delta}{(ig_+-\nu-\Delta)^2-v_F^2k^2}-\frac{ig_+-\nu+\Delta}{(ig_+-\nu+\Delta)^2-v_F^2k^2}\bigg]\nonumber\\
    =&-g_Sg_Bg_-\sigma_z\frac{1}{2}\int \frac{d^2k}{(2\pi)^2} \frac{(\nu-ig_+)^2+(v_F^2k^2-\Delta^2)}{(\nu^2+\Delta^2-g_+^2-v_F^2k^2-2ig_+\nu)^2-4\Delta^2(\nu-ig_+)^2}\nonumber\\
    =&-g_Sg_Bg_-\sigma_z\frac{1}{2}\int \frac{dk}{2\pi} k\frac{(\nu-ig_+)^2+(v_F^2k^2-\Delta^2)}{(\nu^2+\Delta^2-g_+^2-v_F^2k^2-2ig_+\nu)^2-4\Delta^2(\nu-ig_+)^2}.
\end{align}
One can Taylor expand this to linear order in $\nu$ and integrate the result. The primitive function is complicated, and is UV divergent but IR finite. We introduce a UV cutoff $\Lambda$ and then take $m=0$ and $c\rightarrow 0^+$, obtaining that the third diagram in Fig. \ref{fig:SigmaA} evaluates to
\begin{eqnarray}
    -\frac{g_Sg_Bg_-\sigma_z}{8\pi}\bigg[-\frac{2\Lambda^2}{g_+^2+v_F^2\Lambda^2}+\frac{1}{v_F^2}\ln\bigg(1+\frac{v_F^2\Lambda^2}{g_+^2}\bigg)+\frac{2i\Lambda^2(g_+^2-v_F^2\Lambda^2)}{g_+(g_+^2+v_F^2\Lambda^2)^2}\nu\bigg]+\mathcal{O}(\nu^2)
\end{eqnarray}
We note that this is proportional to the result from the previous subsection.

\subsubsection{Adding the terms}
Adding the terms from the two preceding subsections (with a factor of $2$ for the first result, since there are two diagrams there), we have the total advanced self-energy component. 
In the limit $m=0,c\rightarrow 0^+$, we obtain
\begin{eqnarray}\label{sigmares}
    \Sigma^{A}=-\frac{g_Sg_Bg_-\sigma_z}{4\pi}\bigg[-\frac{2\Lambda^2}{g_+^2+v_F^2\Lambda^2}+\frac{1}{v_F^2}\ln\bigg(1+\frac{v_F^2\Lambda^2}{g_+^2}\bigg)+\frac{2i\Lambda^2(g_+^2-v_F^2\Lambda^2)}{g_+(g_+^2+v_F^2\Lambda^2)^2}\nu\bigg]+\mathcal{O}(\nu^2)
\end{eqnarray}
We can also write this as 
\begin{eqnarray}
    \Sigma^{A}=(\Sigma_r-i\Sigma_i)\sigma_z.
\end{eqnarray}
\subsection{Retarded self-energy component}\label{sec:SR}
One can verify that the actions \eqref{SB'} and \eqref{S_H} lead to three Feynman diagrams which are the Hermitian adjoints of those for $\Sigma^A$; see Fig. \ref{fig:SigmaR}.
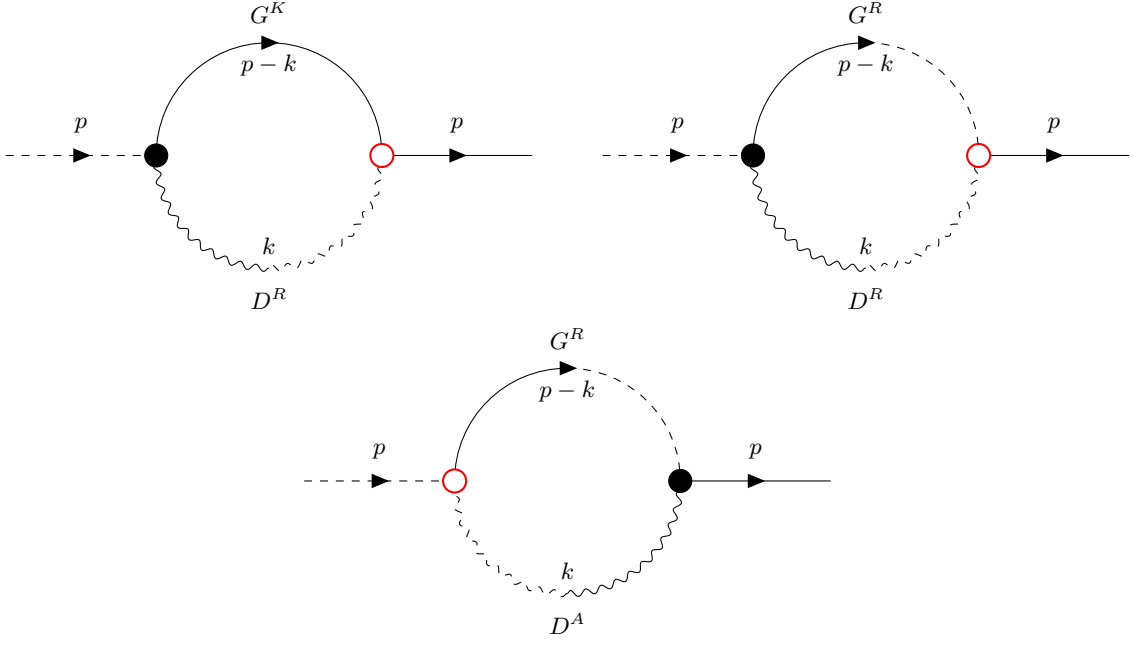
\begin{figure}
    \centering
\begin{tikzpicture}
\begin{feynman}
  \vertex (a) at (0,2);
  \vertex (x) at (2,2);
  \vertex (y) at (5,2);
  \vertex (mid) at (3.5,3.5);
  \vertex (z) at (3.5,0.5);
  \vertex (b) at (7,2);

  \diagram*{
    (a) -- [plain,dashed,with arrow=0.5] (x),
    (x) -- [plain,bend left=45] (mid),
    (mid) -- [plain, with arrow=0,label=above:\(x\),bend left=45] (y),
    (y) -- [plain,with arrow=0.5] (b)
  };
  \diagram*{
    (y) -- [boson ,dashed,bend left=45] (z),
    (z) -- [boson,bend left=45] (x)
  };
\node at ({1,2.4}) {$p$};
\node at ({6,2.4}) {$p$};
\node at ({3.5,3.9}) {$G^K$};
\node at ({3.5,3.2}) {$p-k$};
\node at ({3.5,0.8}) {$k$};
\node at ({3.5,0.1}) {$D^{R}$};
\node [circle,black,draw=black,fill=black] (e) at (2,2);
\node [circle,red,draw=red,thick,fill=white] (f) at (5,2);


\end{feynman}
\end{tikzpicture}
\hspace{20 pt}
\begin{tikzpicture}
\begin{feynman}
  \vertex (a) at (0,2);
  \vertex (x) at (2,2);
  \vertex (y) at (5,2);
  \vertex (mid) at (3.5,3.5);
  \vertex (z) at (3.5,0.5);
  \vertex (b) at (7,2);

  \diagram*{
    (a) -- [plain,dashed,with arrow=0.5] (x),
    (x) -- [plain,bend left=45] (mid),
    (mid) -- [plain,dashed,with arrow=0,label=above:\(x\),bend left=45] (y),
    (y) -- [plain,with arrow=0.5] (b)
  };
  \diagram*{
    (y) -- [boson ,dashed,bend left=45] (z),
    (z) -- [boson,bend left=45] (x)
  };
\node at ({1,2.4}) {$p$};
\node at ({6,2.4}) {$p$};
\node at ({3.5,3.2}) {$p-k$};
\node at ({3.5,3.9}) {$G^{R}$};
\node at ({3.5,0.8}) {$k$};
\node at ({3.5,0.1}) {$D^{R}$};
\node [circle,black,draw=black,fill=black] (e) at (2,2);
\node [circle,red,draw=red,thick,fill=white] (f) at (5,2);


\end{feynman}
\end{tikzpicture}
\begin{tikzpicture}
\begin{feynman}
  \vertex (a) at (0,2);
  \vertex (x) at (2,2);
  \vertex (y) at (5,2);
  \vertex (mid) at (3.5,3.5);
  \vertex (z) at (3.5,0.5);
  \vertex (b) at (7,2);

  \diagram*{
    (a) -- [plain,dashed,with arrow=0.5] (x),
    (x) -- [plain,bend left=45] (mid),
    (mid) -- [plain,dashed, with arrow=0,label=above:\(x\),bend left=45] (y),
    (y) -- [plain, with arrow=0.5] (b)
  };
  \diagram*{
    (y) -- [boson,bend left=45] (z),
    (z) -- [boson,dashed,bend left=45] (x)
  };
\node at ({1,2.4}) {$p$};
\node at ({6,2.4}) {$p$};
\node at ({3.5,3.2}) {$p-k$};
\node at ({3.5,3.9}) {$G^{R}$};
\node at ({3.5,0.8}) {$k$};
\node at ({3.5,0.1}) {$D^{A}$};
\node [circle,red,draw=red,thick,fill=white] (e) at (2,2);
\node [circle,black,draw=black,fill=black] (f) at (5,2);


\end{feynman}
\end{tikzpicture}
\caption{The chirality-mixing diagrams for the self-energy component $\Sigma^{R}$, to one-loop order.}
\label{fig:SigmaR}
\end{figure}
Thus, we immediately have that $\Sigma^{R}=(\Sigma^A)^*$, as it should be. The result is
\begin{eqnarray}\label{sigmares2}
    -\frac{g_Sg_Bg_-\sigma_z}{4\pi}\bigg[-\frac{2\Lambda^2}{g_+^2+v_F^2\Lambda^2}+\frac{1}{v_F^2}\ln\bigg(1+\frac{v_F^2\Lambda^2}{g_+^2}\bigg)-\frac{2i\Lambda^2(g_+^2-v_F^2\Lambda^2)}{g_+(g_+^2+v_F^2\Lambda^2)^2}\nu\bigg]+\mathcal{O}(\nu^2).
\end{eqnarray}
We write this more succinctly as
\begin{eqnarray}
    \Sigma^{R}=(\Sigma_r+i\Sigma_i)\sigma_z.
\end{eqnarray}
\subsection{Keldysh self-energy component}\label{sec:SK}
The diagrams are those in Fig. \ref{fig:SigmaK}.
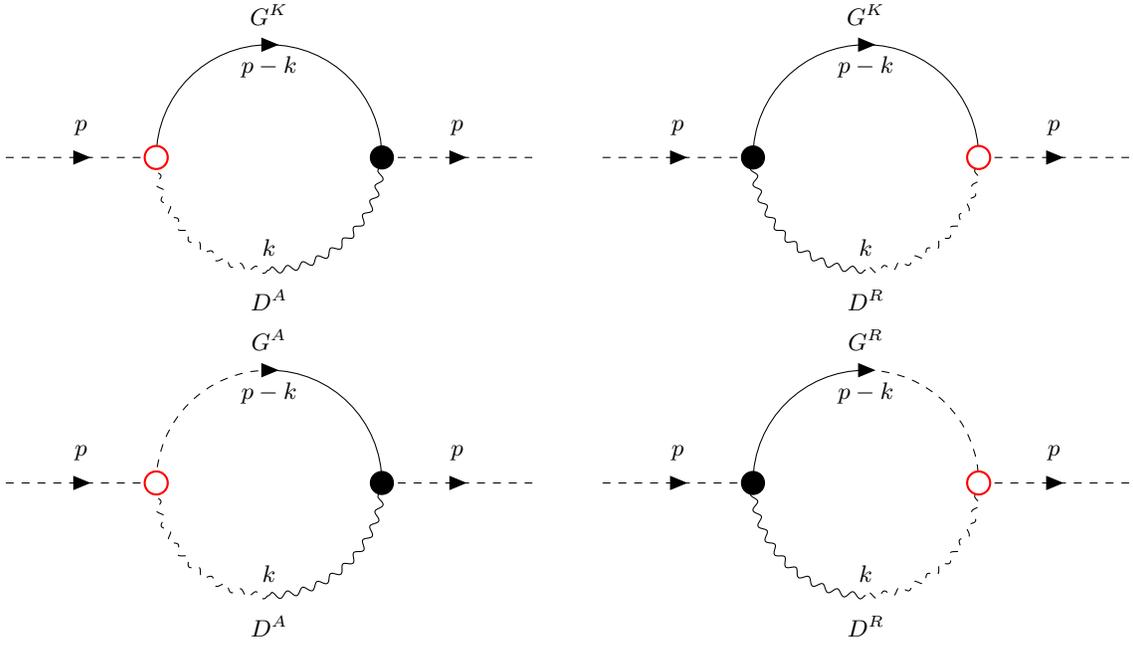
\begin{figure}
    \centering
\begin{tikzpicture}
\begin{feynman}
  \vertex (a) at (0,2);
  \vertex (x) at (2,2);
  \vertex (y) at (5,2);
  \vertex (mid) at (3.5,3.5);
  \vertex (z) at (3.5,0.5);
  \vertex (b) at (7,2);

  \diagram*{
    (a) -- [plain,dashed,with arrow=0.5] (x),
    (x) -- [plain,bend left=45] (mid),
    (mid) -- [plain, with arrow=0,label=above:\(x\),bend left=45] (y),
    (y) -- [plain,dashed,with arrow=0.5] (b)
  };
  \diagram*{
    (y) -- [boson ,bend left=45] (z),
    (z) -- [boson,dashed,bend left=45] (x)
  };
\node at ({1,2.4}) {$p$};
\node at ({6,2.4}) {$p$};
\node at ({3.5,3.9}) {$G^K$};
\node at ({3.5,3.2}) {$p-k$};
\node at ({3.5,0.8}) {$k$};
\node at ({3.5,0.1}) {$D^{A}$};
\node [circle,red,draw=red,thick,fill=white] (e) at (2,2);
\node [circle,black,draw=black,fill=black] (f) at (5,2);


\end{feynman}
\end{tikzpicture}
\hspace{20 pt}
\begin{tikzpicture}
\begin{feynman}
  \vertex (a) at (0,2);
  \vertex (x) at (2,2);
  \vertex (y) at (5,2);
  \vertex (mid) at (3.5,3.5);
  \vertex (z) at (3.5,0.5);
  \vertex (b) at (7,2);

  \diagram*{
    (a) -- [plain,dashed,with arrow=0.5] (x),
    (x) -- [plain,bend left=45] (mid),
    (mid) -- [plain,with arrow=0,label=above:\(x\),bend left=45] (y),
    (y) -- [plain,dashed,with arrow=0.5] (b)
  };
  \diagram*{
    (y) -- [boson ,dashed,bend left=45] (z),
    (z) -- [boson,bend left=45] (x)
  };
\node at ({1,2.4}) {$p$};
\node at ({6,2.4}) {$p$};
\node at ({3.5,3.2}) {$p-k$};
\node at ({3.5,3.9}) {$G^{K}$};
\node at ({3.5,0.8}) {$k$};
\node at ({3.5,0.1}) {$D^{R}$};
\node [circle,black,draw=black,fill=black] (e) at (2,2);
\node [circle,red,draw=red,thick,fill=white] (f) at (5,2);


\end{feynman}
\end{tikzpicture}
\begin{tikzpicture}
\begin{feynman}
  \vertex (a) at (0,2);
  \vertex (x) at (2,2);
  \vertex (y) at (5,2);
  \vertex (mid) at (3.5,3.5);
  \vertex (z) at (3.5,0.5);
  \vertex (b) at (7,2);

  \diagram*{
    (a) -- [plain,dashed,with arrow=0.5] (x),
    (x) -- [plain,dashed,bend left=45] (mid),
    (mid) -- [plain, with arrow=0,label=above:\(x\),bend left=45] (y),
    (y) -- [plain, dashed,with arrow=0.5] (b)
  };
  \diagram*{
    (y) -- [boson,bend left=45] (z),
    (z) -- [boson,dashed,bend left=45] (x)
  };
\node at ({1,2.4}) {$p$};
\node at ({6,2.4}) {$p$};
\node at ({3.5,3.2}) {$p-k$};
\node at ({3.5,3.9}) {$G^{A}$};
\node at ({3.5,0.8}) {$k$};
\node at ({3.5,0.1}) {$D^{A}$};
\node [circle,red,draw=red,thick,fill=white] (e) at (2,2);
\node [circle,black,draw=black,fill=black] (f) at (5,2);


\end{feynman}
\end{tikzpicture}
\hspace{20 pt}
\begin{tikzpicture}
\begin{feynman}
  \vertex (a) at (0,2);
  \vertex (x) at (2,2);
  \vertex (y) at (5,2);
  \vertex (mid) at (3.5,3.5);
  \vertex (z) at (3.5,0.5);
  \vertex (b) at (7,2);

  \diagram*{
    (a) -- [plain,dashed,with arrow=0.5] (x),
    (x) -- [plain,bend left=45] (mid),
    (mid) -- [plain,dashed, with arrow=0,label=above:\(x\),bend left=45] (y),
    (y) -- [plain,dashed, with arrow=0.5] (b)
  };
  \diagram*{
    (y) -- [boson,dashed,bend left=45] (z),
    (z) -- [boson,bend left=45] (x)
  };
\node at ({1,2.4}) {$p$};
\node at ({6,2.4}) {$p$};
\node at ({3.5,3.2}) {$p-k$};
\node at ({3.5,3.9}) {$G^{R}$};
\node at ({3.5,0.8}) {$k$};
\node at ({3.5,0.1}) {$D^{R}$};
\node [circle,black,draw=black,fill=black] (e) at (2,2);
\node [circle,red,draw=red,thick,fill=white] (f) at (5,2);


\end{feynman}
\end{tikzpicture}
\caption{The chirality-mixing diagrams for the self-energy component $\Sigma^{K}$, to one-loop order.}
\label{fig:SigmaK}
\end{figure}
We note that the sum of these diagrams is anti-Hermitian, so $\Sigma^K$ is also manifestly anti-Hermitian, as it should be. We give results for all the diagrams below.
\subsubsection{Upper left diagram in Fig. \ref{fig:SigmaK}}
We have already computed 
\begin{eqnarray}
    ig_Sg_Bg_+\sigma_z\int \frac{d\omega}{2\pi}\frac{d^2k}{(2\pi)^2} D^{A}(\nu-\omega,p-k)G^{K}(\omega,k),
\end{eqnarray}
meaning that our result here is $-\frac{g_-}{g_+}$ times that result. We obtain
\begin{eqnarray}\label{res1}
    \frac{g_Sg_Bg_-^2\sigma_z}{8\pi g_+}\bigg[-\frac{2\Lambda^2}{g_+^2+v_F^2\Lambda^2}+\frac{1}{v_F^2}\ln\bigg(1+\frac{v_F^2\Lambda^2}{g_+^2}\bigg)+\frac{2i\Lambda^2(g_+^2-v_F^2\Lambda^2)}{g_+(g_+^2+v_F^2\Lambda^2)^2}\nu\bigg]+\mathcal{O}(\nu^2)=-\frac{1}{2}\frac{g_-}{g_+}\big(\Sigma_r-i\Sigma_i\big)+\mathcal{O}(\nu^2).
\end{eqnarray}
\subsubsection{Upper right diagram in Fig. \ref{fig:SigmaK}}
This is minus the Hermitian adjoint of the previous result, i.e.
\begin{eqnarray}\label{res2}
    -\frac{g_Sg_Bg_-^2\sigma_z}{8\pi g_+}\bigg[-\frac{2\Lambda^2}{g_+^2+v_F^2\Lambda^2}+\frac{1}{v_F^2}\ln\bigg(1+\frac{v_F^2\Lambda^2}{g_+^2}\bigg)-\frac{2i\Lambda^2(g_+^2-v_F^2\Lambda^2)}{g_+(g_+^2+v_F^2\Lambda^2)^2}\nu\bigg]+\mathcal{O}(\nu^2)=\frac{1}{2}\frac{g_-}{g_+}\big(\Sigma_r+i\Sigma_i\big)+\mathcal{O}(\nu^2)
\end{eqnarray}
\subsubsection{Lower row of diagrams in Fig. \ref{fig:SigmaK}}
These diagrams are minus each other's Hermitian adjoints, and are given by
\begin{eqnarray}
     &-ig_Sg_Bg_+\sigma_z\int \frac{d\omega}{2\pi}\frac{d^2k}{(2\pi)^2} D^{R/A}(\omega,k)G^{R/A}(\nu-\omega,p-k)\nonumber\\&=-ig_Sg_+g_B\sigma_z\int \frac{d\omega}{2\pi}\frac{d^2k}{(2\pi)^2} \frac{1}{2}\frac{1}{\left(\omega\pm i \eta\right)^2-(c_s^2 k^2+m^2)}\frac{1}{(\nu-\omega\pm  ig_+)^2-v_F^2(p-k)^2}\bigg((\nu-\omega \pm ig_+)I+v_F(\vec{p}-\vec{k})\cdot \vec{\sigma}\bigg).
\end{eqnarray}
Only the identity part gives us the desired $\sigma_z$ prefactor. Keeping only that, and setting $p=0$, we have
\begin{eqnarray}
-ig_Sg_Bg_+\sigma_z\int \frac{d\omega}{2\pi}\frac{d^2k}{(2\pi)^2} \frac{1}{2}\frac{1}{\left(\omega\pm i \eta\right)^2-(c_s^2 k^2+m^2)}\frac{1}{(\nu-\omega\pm ig_+)^2-v_F^2k^2}(\nu-\omega\pm ig_+).
\end{eqnarray}
We let $\Delta:=\sqrt{c_s^2k^2+m^2}$ and write
\begin{align}
    \frac{1}{(\omega\pm i\eta)^2-\Delta^2}&=\bigg(\frac{1}{\omega\pm i\eta-\Delta}-\frac{1}{\omega \pm i\eta + \Delta}\bigg)\frac{1}{2\Delta}\nonumber\\
    &=\mp \frac{i\pi}{2\Delta} \big(\delta(\omega-\Delta)-\delta(\omega+\Delta)\big)+\mathcal{P}_{\omega=\Delta}\bigg(\frac{1}{2\Delta(\omega-\Delta)}\bigg)-\mathcal{P}_{\omega=-\Delta}\bigg(\frac{1}{2\Delta(\omega+\Delta)}\bigg),
\end{align}
Although we add the diagrams as $D^RG^{R}+D^{A}G^{A}$, and the delta terms only differ in the overall sign, their contributions do not a priori cancel here since the fermion factors differ. We will therefore need to compute the Dirac delta terms, which are given by
\begin{align}
-ig_Sg_Bg_+\sigma_z&\int \frac{d\omega}{2\pi}\frac{d^2k}{(2\pi)^2} \frac{1}{2}\bigg(\mp \frac{i\pi}{2\Delta}\bigg)\big(\delta(\omega-\Delta)-\delta(\omega+\Delta)\big)\frac{1}{(\nu-\omega\pm ig_+)^2-v_F^2k^2}(\nu-\omega\pm ig_+)\nonumber\\
&=\mp \frac{g_Sg_Bg_+\sigma_z}{8}\int d\omega\frac{d^2k}{(2\pi)^2} \frac{1}{\Delta}\big(\delta(\omega-\Delta)-\delta(\omega+\Delta)\big)\frac{1}{(\nu-\omega\pm ig_+)^2-v_F^2k^2}(\nu-\omega\pm ig_+)\nonumber\\
&=\mp \frac{g_Sg_Bg_+\sigma_z}{8}\int\frac{d^2k}{(2\pi)^2} \frac{1}{\Delta}\bigg(\frac{\nu-\Delta\pm ig_+}{(\nu-\Delta\pm ig_+)^2-v_F^2k^2}-\frac{\nu+\Delta\pm ig_+}{(\nu+\Delta\pm ig_+)^2-v_F^2k^2}\bigg)\nonumber\\
&=\mp \frac{g_Sg_Bg_+\sigma_z}{16 \pi}\int_0^{\infty} kdk \frac{1}{\Delta}\bigg(\frac{\nu-\Delta\pm ig_+}{(\nu-\Delta\pm ig_+)^2-v_F^2k^2}-\frac{\nu+\Delta\pm ig_+}{(\nu+\Delta\pm ig_+)^2-v_F^2k^2}\bigg).
\end{align}
We perform a first-order expansion in $\nu$ and integrate. Adding the four integrals (two for $G^RD^R$, two for $G^{A}D^{A}$) with appropriate signs, we obtain the following total contribution for $c\rightarrow 0^+$:
\begin{eqnarray}\label{resa}
    \frac{g_Sg_B\sigma_z}{4 \pi}\frac{i\Lambda^2(g_+^2-v_F^2\Lambda^2)\nu}{(g_+^2+v_F^2\Lambda^2)^2}.
\end{eqnarray}
For the principal value terms, we use that we know the result \eqref{GADAres}, which is for $ig_Sg_Bg_-\sigma_zG^{A}D^{A}$ and is based purely on the principal value terms from the Sokhotsky-Plemelj theorem. The principal value parts of $-ig_Sg_Bg_+\sigma_zG^{A}D^{A}$ thus become
 \begin{eqnarray}\label{resb}
     \frac{1}{16\pi}g_Sg_Bg_+\sigma_z\bigg[-\frac{2\Lambda^2}{g_+^2+v_F^2\Lambda^2}+\frac{1}{v_F^2}\log\bigg(1+\frac{v_F^2\Lambda^2}{g_+^2}\bigg)+\frac{2i\Lambda^2(g_+^2-v_F^2\Lambda^2)}{g_+(g_+^2+v_F^2\Lambda^2)^2}\nu\bigg]+\mathcal{O}(\nu^2).
 \end{eqnarray}
The principal value part of $-ig_Sg_+\sigma_zG^{R}D^{R}$ is minus the adjoint of eq. \eqref{resb} i.e.
 \begin{eqnarray}\label{resc}
     -\frac{1}{16\pi}g_Sg_Bg_+\sigma_z\bigg[-\frac{2\Lambda^2}{g_+^2+v_F^2\Lambda^2}+\frac{1}{v_F^2}\log\bigg(1+\frac{v_F^2\Lambda^2}{g_+^2}\bigg)-\frac{2i\Lambda^2(g_+^2-v_F^2\Lambda^2)}{g_+(g_+^2+v_F^2\Lambda^2)^2}\nu\bigg]+\mathcal{O}(\nu^2).
 \end{eqnarray}
 Adding up the contributions in \eqref{resa}, \eqref{resb} and \eqref{resc} gives a total contribution
\begin{eqnarray}\label{res3}
    \frac{1}{2\pi}g_Sg_B\sigma_z\frac{i\Lambda^2(g_+^2-v_F^2\Lambda^2)}{(g_+^2+v_F^2\Lambda^2)^2}\nu=i\frac{g_+}{g_-}\Sigma_i
\end{eqnarray}
 for these two diagrams.
 \subsubsection{Adding all the terms together}
The sum of results \eqref{res1}, \eqref{res2} and \eqref{res3} is 
\begin{align}\label{SigmaKapp}
    \Sigma^K=\frac{1}{2\pi}g_Sg_B\bigg[1+\frac{g_-^2}{g^2_+}
    \bigg]\sigma_z\frac{i\Lambda^2(g_+^2-v_F^2\Lambda^2)}{(g_+^2+v_F^2\Lambda^2)^2}\nu=\bigg[\frac{g_+}{g_-}+\frac{g_-}{g_+}
    \bigg]\sigma_zi\Sigma_i\equiv \chi\times \sigma_zi\Sigma_i,
\end{align}
where we have defined the parameter $\chi$.
\section{The polarization tensor}\label{sec:poltensApp}
Let us derive the polarization tensor, with more detail than in the main text.
This is done based on the expression in \eqref{poltens}, or equivalently, the Feynman diagrams shown in Fig. \ref{fig:pol}. Our propagators are the dressed propagators discussed in Sec. \ref{sec:pol}, with the retarded and advanced components given by eq. \eqref{GRA}. The Keldysh component is
 \begin{eqnarray}\label{GKtot}
     G^K=G^K_{\chi}+G^K_{-},
 \end{eqnarray}
 where the terms are given by
\begin{align}\label{GKchi}
    G^K_{\chi}&=i\chi \Sigma_i\frac{2(g_+\Sigma_i+\Sigma_r\omega)I+
    2v_F(g_+k_y+k_x\Sigma_r)\sigma_x-2v_F(g_+k_x-k_y\Sigma_r)\sigma_y+(g_+^2+\Sigma_i^2+\Sigma_r^2-v_F^2k^2+\omega^2)\sigma_z}{g_+^4+\Sigma_i^4-8g_+\Sigma_i\Sigma_r\omega+(\Sigma_r^2+v_F^2k^2-\omega^2)^2+2\Sigma_i^2(\Sigma_r^2-v_F^2k^2+\omega^2)+2g_+^2(-\Sigma_i^2+\Sigma_r^2+v_F^2k^2+\omega^2)}
\end{align}
and
\begin{align}\label{GK-}
    G^K_{-}&= 2ig_-\frac{(g_+^2+\Sigma_i^2+\Sigma_r^2+v_F^2k^2+\omega^2)I+
    2v_F(k_y\Sigma_i+k_x\omega)\sigma_x-2v_F(k_x\Sigma_i-k_y\omega)\sigma_y+2(g_+\Sigma_i+\Sigma_r\omega)\sigma_z}{g_+^4+\Sigma_i^4-8g_+\Sigma_i\Sigma_r\omega+(\Sigma_r^2+v_F^2k^2-\omega^2)^2+2\Sigma_i^2(\Sigma_r^2-v_F^2k^2+\omega^2)+2g_+^2(-\Sigma_i^2+\Sigma_r^2+v_F^2k^2+\omega^2)}.
\end{align}

We compute the contributions of these two diagrams, and find a polarization tensor which is on the form suitable for a Chern-Simons term. 
Neglecting any renormalization of the electron charge $e$, we have that our first Feynman diagram contribution is
\begin{eqnarray}
    -\frac{i}{2}e^2\mathrm{Tr}\int  \frac{d\omega}{2\pi}\int \frac{d^2k}{(2\pi)^2}\gamma_{\mu}G^R(\nu-\omega,p-k)\gamma_{\nu}G^K(\omega,k).
\end{eqnarray}
Since $\mu,\nu\in\{1,2\}$ for the elements of the polarization tensor we consider, there has to be one $\gamma_0=\sigma_z$ factor and an identity matrix factor from the propagators to produce a nonzero trace; c.f. \eqref{minkowski}. This means that there are only two integrals to compute for the $G^RG^K$ part. Similar reasoning applies to the second diagram in Fig. \ref{fig:pol}, yielding four terms in total. We compute them with $\Sigma_i(\omega)= \omega\bar{\Sigma}_i$ to make the frequency dependence explicit.

We take the self-energies from before, as well as the $I$ and $\sigma_z$ ($\sigma_z$ and $I$) terms of $G^R$ ($G^K$). We then gather the matrix factors as $\gamma_{\mu}I\gamma_{\nu}\sigma_z=\gamma_{\mu}\gamma_{\nu}\sigma_z$ and $\gamma_{\mu}\sigma_z\gamma_{\nu} I=-\gamma_{\mu}\gamma_{\nu}\sigma_z$ for the values of interest of the index $\nu$, only making the following approximation: we series expand in $\nu$ around $\nu=p_x=p_y=0$ and take the $\nu^1$ term. This means that for that term, the external momentum has been set to $0$ in what follows. We then rescale the variables by instead using 
\begin{eqnarray}
    w=\frac{\omega}{g_+},\quad \kappa=\frac{v_F k}{g_+},\quad \sigma_r=\frac{\Sigma_r}{g_+},
\end{eqnarray}
where the frequency-independent $\bar{\Sigma}_i$ is not rescaled because the rescaling of $\Sigma_i(\omega)=\omega\bar{\Sigma}_i$ is taken care of by the frequency rescaling.
However, we do not rescale the overall factor $\nu$. The measure rescales as
\begin{eqnarray}
    \frac{d\omega}{2\pi}\frac{d^2k}{(2\pi)^2}\rightarrow \frac{g_+^3}{v_F^2}\frac{dw}{2\pi}\frac{d^2\kappa}{(2\pi)^2},
\end{eqnarray}
and we also rescale the self-energy momentum cutoff as the momenta by introducing
\begin{eqnarray}
    \tilde{\Lambda}=\frac{v_F\Lambda}{g_+}.
\end{eqnarray}
By performing the integrals and inserting our self-energy expressions as well as $\chi$, and then taking the limit of small gain/loss imbalance $g_-$, we obtain
\begin{eqnarray}
    -\frac{ie^2}{2}\times(-2i\epsilon^{0\mu\nu})\times \bigg(-\frac{ig_Bg_S\tilde{\Lambda}^2(\tilde{\Lambda}^2-1)\nu}{128\pi^2v_F^4(\tilde{\Lambda}^2+1)^2}\bigg)\rightarrow ie^2\nu\epsilon^{0\mu\nu}\frac{g_Bg_S}{128\pi^2v_F^4},\quad \tilde{\Lambda}\rightarrow \infty,
\end{eqnarray}
where we have used the identity \eqref{epsilonid} for the middle factor in the left-hand side, and the third factor comes from the integral. 
Similar steps, where only the linear term in $p_x$ or $p_y$ is kept, give an integrand which is odd in either $k_x$ or $k_y$, and hence integrates to $0$. Thus, the above holds more generally with $\nu\epsilon^{0\mu\nu}\rightarrow p_{\rho}\epsilon^{\rho\mu\nu}$ for $\mu,\nu\in\{1,2\}$. All in all, we have the polarization tensor component
\begin{eqnarray}\label{pol1}
   \Pi_{\mu\nu}^{R}=-ie^2p_{\rho}\epsilon^{\mu\rho\nu}\frac{g_Bg_S}{128\pi^2v_F^4}
,\qquad \mu,\nu\in\{1,2\}.
\end{eqnarray}
This polarization tensor immediately leads to the Chern-Simons term, as explained in the main text.
We note that the polarization tensor vanishes in the zero-frequency limit, and that its sign is determined by the fermion-boson couplings of both the jump operators ($g_B$) and the non-dissipative part of the action \eqref{S_H} ($g_S$).

\section{Results neglecting wavefunction renormalization}\label{sec:pol2app}

Neglecting the renormalization of the wavefunction, which is equivalent to performing the calculation for the self-energy in Sec.~\ref{sec:self} with the external frequency equal to zero, the form of the dressed Green functions and the calculation of the polarization tensor simplify considerably since both the imaginary part and the Keldysh component of the self-energy vanish. 

In fact, for zero $\Sigma_i$ and $\Sigma^K$, the Green's functions take the form
\begin{equation}\label{GRA2}
    G^{R/A}(\omega, \vec{k})=\frac{1}{\left(\nu\pm i g_+\right)I-v_F\vec{k}\cdot \vec{\sigma}-\Sigma_r \sigma_z}
\end{equation}
and
\begin{equation}\label{GK2}
    G^K(\omega, \vec{k})=i g_-G^R(\omega, k)G^A(\omega, k)=i g_-\frac{1}{\left(\nu+ i g_+\right)I-v_F\vec{k}\cdot \vec{\sigma}-\Sigma_r \sigma_z}\frac{1}{\left(\nu- i g_+\right)I-v_F\vec{k}\cdot \vec{\sigma}-\Sigma_r \sigma_z}.
\end{equation}

To obtain a Chern-Simons action of the form of Eq.~\eqref{eq_Gamma_CS}, we need to have a contribution of the polarization tensor which looks like
\begin{equation}
    \Pi^{\mu\nu}_{CS}(q)=e^2\frac{k}{4\pi}\epsilon^{\mu\nu\rho}q_\rho.    
\end{equation}
Here, $q_0=\nu$ and $q_{i}=-v_F k_i$ are the external momentum for the polarization. To extract the level $k$ we can use a procedure similar to what is done in the usual calculation of the Chern-Simons level for a massive Dirac fermion \cite{zee2010quantum}: we Taylor expand in external momentum up to first order, and isolate the linear term. This reduces to taking the derivative with respect to $q_\rho$ and then setting the external momentum to zero, after which the general expression is contracted with $\epsilon_{\mu\nu\rho}$. 

The only terms that depend on momentum are the propagators. They have the derivatives
\begin{eqnarray}
    \hspace{-1cm}\left.\partial_{q_\rho}G^{R/A}(\omega+\nu, \vec{p}+\vec{k})\right|_{q=0}&=&-\frac{1}{\left(\omega\pm i g_+\right)I-v_F\vec{p}\cdot \vec{\sigma}-\Sigma_r \sigma_z}\gamma^0\gamma^\rho\frac{1}{\left(\omega\pm i g_+\right)I-v_F\vec{p}\cdot \vec{\sigma}-\Sigma_r \sigma_z}\equiv-\frac{1}{A_{R/A}}\sigma^\rho\frac{1}{A_{R/A}}\\
    \hspace{-1cm}\left.\partial_{q_\rho}G^{K}(\omega+\nu, \vec{p}+\vec{k})\right|_{q=0}&=&ig_-\left[\left.\partial_{q_\rho}G^{R}(\omega+\nu, \vec{p}+\vec{k})G^{A}(\omega+\nu, \vec{p}+\vec{k})\right|_{q=0}+\left.G^{R}(\omega+\nu, \vec{p}+\vec{k})\partial_{q_\rho}G^{A}(\omega+\nu, \vec{p}+\vec{k})\right|_{q=0}\right]\nonumber\\
    \hspace{-1cm}&=&-i g_-\left[\frac{1}{A_R}\sigma^\rho\frac{1}{A_R}\frac{1}{A_A}+\frac{1}{A_R}\frac{1}{A_A}\sigma^\rho\frac{1}{A_A}\right],
\end{eqnarray}
where we introduced the matrices $A_{R/A}=\left(\omega\pm i g_+\right)I-v_F\vec{p}\cdot \vec{\sigma}-\Sigma_r \sigma_z$ and used that $\gamma^0\gamma^\rho=\sigma^\rho=(I, \sigma_x, \sigma_y)$ to reduce the size of the expressions. 

Using Eq.~\eqref{eq:pol_Keldysh}, we can then express the level $k$ as
\begin{equation}
    k=-i \pi g_- \int \frac{d\omega}{2\pi}\int \frac{d^2p}{(2\pi)^2}\, \epsilon_{\mu\nu\rho}\mathrm{Tr}\left[\sigma^{\mu}\left(\frac{1}{A_{R}}\sigma^\rho\frac{1}{A_{R}}\sigma^\nu\frac{1}{A_{R}}\frac{1}{A_{A}}+\frac{1}{A_{R}}\sigma^{\nu}\frac{1}{A_R}\sigma^\rho\frac{1}{A_{R}}\frac{1}{A_{A}}+\frac{1}{A_{R}}\sigma^{\nu}\frac{1}{A_R}\frac{1}{A_{A}}\sigma^\rho\frac{1}{A_{A}}\right)\right].
\end{equation} 
We now write 
\begin{equation}
        \frac{1}{A_{R/A}}=\frac{1}{\left(\omega\pm i g_+\right)I-v_F\vec{p}\cdot \vec{\sigma}-\Sigma_r \sigma_z}=\frac{\left(\omega\pm i g_+\right)I+v_F\vec{p}\cdot \vec{\sigma}+\Sigma_r \sigma_z}{\left(\omega\pm i g_+\right)^2-v_F^2p^2-\Sigma_r^2}=\frac{B_{R/A}}{C_{R/A}},
\end{equation}
such that 
\begin{equation}
        k=-i \pi g_- \int \frac{d\omega}{2\pi}\int \frac{d^2p}{(2\pi)^2}\, \epsilon_{\mu\nu\rho}\mathrm{Tr}\left[\sigma^{\mu}\left(\frac{B_R\sigma^\rho B_R \sigma^\nu B_R B_A}{C_R^3 C_{A}}+\frac{B_R\sigma^\nu B_R\sigma^\rho B_RB_A}{C_R^3 C_{A}}+\frac{B_R\sigma^\nu B_R B_A\sigma^\rho B_A}{C_R^2 C_{A}^2}\right)\right].
\end{equation}
We see that the first two terms cancel, and the only remaining one is 
\begin{equation}
        k=-i \pi g_- \int \frac{d\omega}{2\pi}\int \frac{d^2p}{(2\pi)^2}\, \epsilon_{\mu\nu\rho}\frac{\mathrm{Tr}\left(\sigma^\mu B_R\sigma^\nu B_R B_A\sigma^\rho B_A\right)}{C_R^2 C_{A}^2}.
\end{equation}
In this form, one can compute the trace and obtain
\begin{equation}
   k=-16\pi g_-\Sigma_r \int \frac{d\omega}{2\pi}\int\frac{d^2p}{(2\pi)^2}\frac{\omega(-g_+^2+\Sigma_r^2+v_F^2p^2-\omega^2)}{\left[\left(\omega+ i g_+\right)^2-v_F^2p^2-\Sigma_r^2\right]^2\left[\left(\omega-i g_+\right)^2-v_F^2p^2-\Sigma_r^2\right]^2}.
\end{equation}
Since this is an odd function of frequency, the integral vanishes identically, showing that there is no Hall effect without considering the renormalization of the wavefunction. This should be contrasted to the usual case where a finite mass term leads to the breaking of TRI and, consequently, a finite quantum Hall conductance. 
\end{widetext}
\end{document}